\documentclass[prd,aps,showpacs,floats,floatfix,groupedaddress,letterpaper]{revtex4}
\usepackage{amssymb,amsmath}
\usepackage{dcolumn,epsfig}
\usepackage{amsfonts}
\usepackage{latexsym}
\usepackage{amssymb}
\usepackage{hyperref}
\usepackage{graphicx}
\usepackage{relsize}
\usepackage{anyfontsize}
\usepackage{mathtools}

\newcommand{\be}{\begin{equation}}
\newcommand{\ee}{\end{equation}}
\newcommand{\beq}{\begin{eqnarray}}
\newcommand{\eeq}{\end{eqnarray}}
\newcommand{\bed}{\begin{displaymath}}
\newcommand{\eed}{\end{displaymath}}
\newcommand{\bc}{\begin{center}}
\newcommand{\ec}{\end{center}}
\newcommand{\bi}{\begin{itemize}}
\newcommand{\ei}{\end{itemize}}
\newcommand{\bn}{\begin{enumerate}}
\newcommand{\en}{\end{enumerate}}

\begin{document}

\title{Invariant quantities in the scalar-tensor theories of gravitation}

\author{Laur J\"arv} \email{laur.jarv@ut.ee}
\affiliation{Institute of Physics, University of Tartu,
Ravila 14c, Tartu 50411, Estonia}
\author{Piret Kuusk} \email{piret.kuusk@ut.ee} 
\affiliation{Institute of Physics, University of Tartu,
Ravila 14c, Tartu 50411, Estonia}
\author{Margus Saal}\email{margus.saal@ut.ee}
\affiliation{Institute of Physics, University of Tartu,
Ravila 14c, Tartu 50411, Estonia}
\author{Ott Vilson}\email{ovilson@ut.ee}
\affiliation{Institute of Physics, University of Tartu,
Ravila 14c, Tartu 50411, Estonia}
\date{\today}


\begin{abstract}
We consider the general scalar-tensor gravity without derivative couplings. By rescaling of the metric and reparametrization of the scalar field, the theory can be presented in different conformal frames and parametrizations. In this work we argue, that while due to the freedom to transform the metric and the scalar field, the scalar field itself does not carry a physical meaning (in a generic parametrization), there are functions of the scalar field and its derivatives which remain invariant under the transformations. We put forward a scheme how to construct these invariants, discuss how to formulate the theory in terms of the invariants, and show how the observables like parametrized post-Newtonian parameters and characteristics of the cosmological solutions can be neatly expressed in terms of the invariants. In particular, we describe the scalar field solutions in Friedmann-Lema\^itre-Robertson-Walker cosmology in Einstein and Jordan frames, and explain their correspondence despite the approximate equations turning out to be linear and non-linear in different frames.
\end{abstract}

\maketitle


\section{Introduction}


Scalar-tensor gravity (STG) \cite{jordanfierz, BDBW, stg_books, Flanagan} introduces a scalar field that is nonminimally coupled to curvature and thus can be interpreted as an additional mediator of gravitational interaction besides the usual metric tensor.
Such theories provide a simple but versatile extension to general relativity, 
they arise naturally in constructions involving higher dimensions,
and feature in attempts to construct scale-invariant fundamental physics \cite{deser}.
The theory can be generalized further by allowing various derivative couplings and higher order derivative terms in the action \cite{Horndeski}. It has received a lot of attention
in phenomenological model building: inflation and dark energy \cite{dark_energy}, and more recently Higgs inflation \cite{Higgs_inflation}.

Since the early paper by Dicke \cite{dicke} it has been well known that by rescaling of the metric and reparametrization of the scalar field, the theory can be presented in different conformal frames and parametrizations \cite{Flanagan}.
Despite an extensive use of this property as a convenient
calculational tool, there lingers a conceptual issue of what is the precise relation of different frames and parametrizations 
 to the observable world and to each other.
 
In the former aspect it is a question whether physical measurements choose one frame which defines the units used in physical observations, i.e.~which metric defines the measured lengths (for early references see Refs. \cite{stg_books,Faraoni:1998qx} while some recent papers are Refs. \cite{physical_equivalence}). 
From an alternative point of view, also letting the units to rescale inversely with the metric neutralizes the effect of conformal transformation \cite{dicke,faraoni_units}, and the question of physical frame becomes superfluous. This can be interpreted by generalizing the underlying geometry from Riemann into Weyl-integrable  \cite{quiros}.

The latter aspect means a mathematical problem, whether the different formulations are mathematically equivalent. Here the common wisdom about the subject says that different frames are equivalent on the level of classical action (although one must be careful in the limit where the transformation becomes singular \cite{JKS_Einstein_Jordan}). However, things get more complicated and warrant a careful consideration and debate on the level of e.g.\ cosmological perturbations \cite{cosmological_perturbations} and quantum corrections \cite{quantum_corrections,Kamenshchik:2014waa}.

One may view different conformal frames and parametrizations of the theory as arising from a change of coordinates in some abstract generalized field space. Then the discrepancies can be attributed to the circumstance that the theory has not been formulated in a covariant way with respect to that abstract space \cite{Kamenshchik:2014waa}.
Therefore some authors have strived to formulate the theory in terms of invariant variables.
The idea has been to focus upon the conformal transformation and express all observables in terms of frame-invariant combinations of the theory parameters and variables, as well as the units \cite{catena,postma}.

In the present paper we complement this line of thought by introducing invariant quantities of the scalar field.
The scalar field is amenable to reparametrization, therefore in a generic parametrization it can not carry a physical meaning (can not be measured directly). 
However, it is possible to combine the functions of scalar field and their derivatives into quantities which remain invariant under the conformal transformations and field redefinitions, and therefore should have a more direct relevance to observable physics. Indeed, using these quantities we show how the parametrized post-Newtonian (PPN) parameters and the qualitative features of the scalar field cosmological solutions like convergence properties and periods of oscillation are independent of the frame and parametrization. These invariant quantities also enable us to write the equations of motion and the action in a manifestly invariant form, and ease the conversion of calculational results 
from one frame and parametrization into another.
A few preliminary efforts in this approach were presented earlier in a conference note \cite{JKSV_1}.

The outline of the paper is as follows. First we recall the general action for scalar-tensor gravity and 
the rules of transformation under conformal rescaling and field reparametrization.
In the next section we introduce three basic invariant quantities of the scalar field and outline how to construct many other invariants from them.
In Sec.~4 we invoke an invariant metric that helps to write the field equations and the action in terms of the invariants. As an application in Sec.~5 we convert the PPN parameters into an invariant form and check that they reproduce the results for particular parametrizations known in the literature. 
In Sec.~6 we focus upon the flat Friedmann-Lema\^itre-Robertson-Walker (FLRW) universe without matter, and study the scalar field solutions near the fixed points.
The conditions for the fixed points as well as the eigenvalues determining the approximate solutions turn out to be invariant. 
Yet for a specific situation it is interesting to see, how a linear result in the Einstein frame
can actually correspond to a nonlinear result in the Jordan frame.
Finally, in Sec.~7 we conclude with a brief summary and outlook.


\section{General action functional and different parametrizations}

\subsection{General action functional}
\label{general action functional}
Let us consider the general action functional for a scalar-tensor theory of gravity 
written down by Flanagan \cite{Flanagan},
\be
\label{fl_moju}
S = \frac{1}{2\kappa^2}\int_{V_4}d^4x\sqrt{-g}\left\lbrace {\mathcal A}(\Phi)R-
{\mathcal B}(\Phi)g^{\mu\nu}\nabla_\mu\Phi \nabla_\nu\Phi - 2\ell^{-2}{\mathcal V}(\Phi)\right\rbrace 
+ S_m\left[e^{2\alpha(\Phi)}g_{\mu\nu},\chi\right] \,.
\ee
It contains four arbitrary functions of the dimensionless scalar field $\Phi$: curvature coupling
function ${\mathcal A}(\Phi)$, generic kinetic coupling of the scalar field ${\mathcal B}(\Phi)$,
self-interaction potential of the scalar field ${\mathcal V}(\Phi)$ and 
conformal coupling $e^{2\alpha(\Phi)}$
between the metric $g_{\mu\nu}$ and matter fields $\chi$. Functions 
${\mathcal A}(\Phi)$, ${\mathcal B}(\Phi)$, ${\mathcal V}(\Phi)$ and
$\alpha(\Phi)$ are dimensionless and fixing them all gives us some concrete theory. 
In the rest of the text we drop the arguments of functions unless confusion might arise.
 
If we impose a physical condition that gravitational interaction is always finite and attractive, the curvature coupling function must satisfy  $0 < \mathcal{A}(\Phi) < \infty$. We also assume from physical considerations  that self-interaction potential is nonnegative,  $0 \leq {\mathcal V}(\Phi) < \infty  $. We will use the units where $c = 1$, but we do not fix the values of the nonvariable part of the effective gravitational ``constant" $\kappa^2$ and a positive constant parameter $\ell$ with the dimension of length, e.g. the Planck length. Note that from a convention $[\kappa^2] = 1 $ it follows that $[S] = [\hbar] = L^2$ and from a convention $[S] = [\hbar] = 1 $ it follows that  $[\kappa^2] = L^2 $.

It is well known that two out of the four arbitrary functions $\left\lbrace \mathcal{A},\,\mathcal{B},\,\mathcal{V},\,\alpha \right\rbrace$ can 
be fixed by transformations that contain two functional degrees of freedom
\beq
\label{conformal_transformation}
	g_{\mu\nu} &=& e^{2\bar{\gamma}(\bar{\Phi})}\bar{g}_{\mu\nu} \,,	 \\
\label{field_redefinition}
	\Phi &=& \bar{f}(\bar{\Phi}) \,.
\eeq
We shall refer to first of them as the change of the frame and the second one the reparametrization of the scalar field. The change of the frame is in fact a conformal rescaling of the metric. We assume that the function $\bar{\gamma}(\bar{\Phi})$ and its first and second derivative, $d\bar{\gamma}/d\bar{\Phi}$ and $d^2\bar{\gamma}/d\bar{\Phi}^2$ respectively, do not diverge at any permitted $\bar{\Phi}$, because otherwise we would introduce geometrical singularities via conformal transformation. 
(Note that this excludes the interesting possibility of ``conformal continuation" \cite{Bronnikov:2002kf}.)
We also assume the function $\bar{f}(\bar{\Phi})$ to be at least directionally continuous, but retain a possibility that Jacobian $\bar{f}^\prime \equiv d\Phi/d\bar{\Phi}$ of this coordinate transformation in $1$-dimensional field space may be singular at some isolated value of the scalar field $\bar{\Phi}$. 

Under the transformation (\ref{conformal_transformation}), (\ref{field_redefinition}) the action functional (\ref{fl_moju}) preserves its structure up to the boundary term (total divergence)
\beq 
\nonumber
 {\bar S} =& \frac{1}{2\kappa^2}\int_{V_4}d^4x\sqrt{-{\bar g}}
 \{ {\bar {\mathcal A}}({\bar \Phi}){\bar R}-{\bar {\mathcal B}}({\bar \Phi})
 {\bar g}^{\mu\nu}{\bar \nabla}_\mu{\bar \Phi}{\bar \nabla}_\nu{\bar \Phi} - 2\ell^{-2}{\bar {\mathcal V}}
 ({\bar \Phi}) \} + \bar{S}_m\left[e^{2{\bar \alpha}({\bar \Phi})}{\bar g}_{\mu\nu},\chi\right] \\
 \label{fl_teisendatud_moju}
 & - \frac{1}{2\kappa^2}\int_{V_4}d^4x \, \partial_\mu \left(6\bar{\gamma}^\prime\sqrt{-\bar{g}}\bar{\mathcal{A}} \, \bar{g}^{\mu\nu}\partial_\nu\bar{\Phi}\right)\,,
\eeq
with transformed functions \cite{Flanagan}
\be
\label{fl_fnide_teisendused}
\begin{array}{rcl}
	\bar{\mathcal{A}}(\bar{\Phi}) &=& e^{2\bar{\gamma}(\bar{\Phi})}
	{\mathcal A} \left( {\bar f}( {\bar \Phi})\right) \,,\\
	{\bar {\mathcal B}}({\bar \Phi}) &=& e^{2{\bar \gamma}({\bar \Phi})}\left( 
	\left(\bar{f}^\prime\right)^2{\mathcal B}\left(\bar{f}(\bar{\Phi})\right) -
	 6\left(\bar{\gamma}^\prime\right)^2{\mathcal A}\left(\bar{f}(\bar{\Phi})\right) -
	  6\bar{\gamma}^\prime\bar{f}^\prime \mathcal{A}^\prime \right) \,, \\
	\bar{{\mathcal V}}(\bar{\Phi}) &=& e^{4\bar{\gamma}(\bar{\Phi})} \, {\mathcal V}\left(\bar{f}(\bar{\Phi})\right) \,, \\
	\bar{\alpha}(\bar{\Phi}) &=& \alpha\left(\bar{f}(\bar{\Phi})\right) + \bar{\gamma}(\bar{\Phi})\,.
\end{array}
\ee
Here we have adopted a convention that prime at a quantity with a bar denotes derivative with respect to 
$\bar{\Phi}$, e.g. $\bar{f}^\prime  \equiv \displaystyle{\frac{d\bar{f}(\bar{\Phi})}{d\bar{\Phi}}}$, and prime at a quantity without a bar
denotes derivative with respect to $\Phi$, e.g.  $\mathcal{A}^\prime \equiv \displaystyle{\frac{d \mathcal{A}(\Phi)}{d \Phi}}$. 
If we denote the backward transformations as
\beq
\label{conformal_transformation_backwards}
\bar{g}_{\mu\nu} &=& e^{2\gamma(\Phi)}g_{\mu\nu}  \,,\\
\label{field_redefinition_backwards}
\bar{\Phi} &=& f(\Phi) \,,
\eeq
then $\gamma\left( \bar{f}\left(\bar{\Phi}\right) \right) = -\bar{\gamma}\left( \bar{\Phi} \right)$.

Under the assumptions on $\bar{\gamma}$ and its derivatives mentioned above, the transformation rules
(\ref{fl_fnide_teisendused}) imply the following.
\begin{itemize}

\item The conditions on curvature coupling function, $0 < {\mathcal{A}}  < \infty $, and 
self-interaction potential, $0 \leq {\mathcal V} < \infty$, are
preserved, i.e.  $0 < {\bar{\mathcal{A}}}  < \infty$ and ${0 \leq \bar {\mathcal V}} < \infty$.

\item If in some frame $\alpha =0 $, then in any other frame $|{\bar\alpha} | < \infty$. 

\item If we want to avoid ghosts, i.e. if there is a frame where the tensorial and scalar part of the gravitational interaction are separated with ${\mathcal A}=1$ and ${\mathcal B}>0$, then in any related frame and parametrization it follows that $2 \bar{\mathcal{A}} \bar{\mathcal{B}} + 3 \left(\bar{\mathcal{A}}^{\prime}\right)^2$ is nonnegative. In this text we assume this quantity to be also nonvanishing. In other words we assume a strict inequality
\be \label{no_ghosts}
\bar{\mathcal{F}} \equiv \frac{2 \bar{\mathcal{A}} \bar{\mathcal{B}} + 3 \left(\bar{\mathcal{A}}^{\prime}\right)^2}{4\bar{\mathcal{A}}^2} > 0 \,.
\ee
However, we do not impose a condition that the quantity $\bar{\mathcal{F}}$ is bounded from above.

\end{itemize}  


\subsection{Different parametrizations}

In the literature mostly such action functionals are considered where two out of the four arbitrary functions $\left\lbrace \mathcal{A},\,\mathcal{B},\,\mathcal{V},\,\alpha \right\rbrace$ are fixed. 
If the latter can be derived from the action functional (\ref{fl_moju}) by using the transformations (\ref{conformal_transformation}) and (\ref{field_redefinition}) then the corresponding theory retains its generality up to some details. 
We use the term `fixed parametrization' to refer to 
the case when two arbitrary functions out of four are fixed by the transformations. 
Fixing the remaining two functions gives a specific theory in this parametrization.
The most common parametrizations are the following.

\begin{itemize}
	\item The Jordan frame action in the Brans-Dicke-Bergmann-Wagoner parametrization (JF BDBW) 
	\cite{BDBW} for the 
	scalar field $\Psi$ fixes ${\mathcal A} = \Psi$, $\alpha = 0$, while keeping 
	${\mathcal B} = \omega(\Psi)/\Psi$, ${\mathcal V}= {\mathcal V}(\Psi)$. 

	\item  The Jordan frame action in the parametrization used by e.g. Boisseau, Esposito-Far\`{e}se,  
	Polarski and Starobinsky (JF BEPS) \cite{BEPS} for the scalar field $\phi$ is obtained by taking ${\mathcal B} = 1$, $\alpha = 0$, while having ${\mathcal A} = F(\phi)$, 
	${\mathcal V}  = {\mathcal V}(\phi)$.

	\item  The Einstein frame action in canonical parametrization (EF canonical) \cite{BDBW, dicke} 
	for the scalar field $\varphi$, fixes
	${\mathcal A} = 1$, ${\mathcal B} = 2$,
	while keeping $\alpha = \alpha(\varphi)$ and ${\mathcal V}={\mathcal V}(\varphi)$. This is the parametrization that was meant when no ghost condition (\ref{no_ghosts}) was discussed. 
\end{itemize}

In the Jordan frame the metric tensor that is used to construct geometrical objects is the same that enters the matter part of the action functional. Therefore freely falling particles follow the geodesics of the corresponding geometry. In the Einstein frame scalar and tensor degrees of freedom are separated and a well-posed initial value formulation is guaranteed by general theorems \cite{Wald}.

\section{Invariants} 
\subsection{Constructing invariants}
 A closer look at the transformations (\ref{fl_fnide_teisendused}) allows us to write out four quantities that under the rescaling (\ref{conformal_transformation}) and reparametrization (\ref{field_redefinition}) gain a multiplier but otherwise preserve their structure. Namely: 
 \beq
 \label{semiinvariant_1}
 \bar{\mathcal{A}} &=& e^{2\bar{\gamma}}\mathcal{A} \,,\\
 \label{semiinvariant_2}
 e^{2\bar{\alpha}} &=& e^{2\bar{\gamma}}e^{2\alpha} \,,\\
 \label{semiinvariant_3}
 \bar{\mathcal{V}} &=& e^{4\bar{\gamma}}\mathcal{V} \,,\\
 \label{semiinvariant_4}
 \bar{\mathcal{F}} \equiv \frac{2\bar{\mathcal{A}}\bar{\mathcal{B}} + 3\left(\bar{\mathcal{A}}^\prime\right)^2}{4\bar{\mathcal{A}}^2} &=& \left(\bar{f}^\prime\right)^2\frac{2\mathcal{AB} + 3\left(\mathcal{A}^\prime\right)^2}{4\mathcal{A}^2} \equiv \left(\bar{f}^\prime\right)^2\mathcal{F} \,.
 \eeq
From these we can construct three independent quantities that are invariant under a local rescaling of the metric tensor and transform as scalar functions under the scalar field redefinition: 
\beq
\label{I_1}
\mathcal{I}_1(\Phi) &\equiv& \frac{e^{2\alpha(\Phi)}}{\mathcal{A}(\Phi)} \,, \\
\label{I_2}
\mathcal{I}_2(\Phi) &\equiv& \frac{\mathcal{V}(\Phi)}{\left(\mathcal{A}(\Phi)\right)^2} \,, \\
\label{I_3}
\mathcal{I}_3(\Phi) &\equiv& \pm\mathop{\mathlarger{\int}}\sqrt{\mathcal{F}(\Phi)}d\Phi \,.
\eeq
Note that at any space-time point $x \in V_4$ the scalar field values are related to each other via Eq.~(\ref{field_redefinition}) and therefore we are actually dealing with space-time point invariants,
\be
\bar{\mathcal{I}}_i (\bar{\Phi}(x)) = \mathcal{I}_i (\bar{f}\left(\bar{\Phi}(x)\right)) = \mathcal{I}_i (\Phi(x)) \,.
\ee
This means that their numerical value at some fixed space-time point is preserved under the transformation (\ref{field_redefinition}) while their functional form with respect to the scalar field as an argument changes under that transformation. We shall refer to these quantities 
as invariants.
As the conformal transformation or the scalar field redefinition are in principle unrelated to a coordinate transformation it follows that spacetime derivatives of invariants,
 \be
 \partial_\mu\left( \bar{\mathcal{I}}_i\left(\bar{\Phi}(x)\right)\right) = \partial_\mu\left( \mathcal{I}_i\left(\bar{f}\left(\bar{\Phi}(x)\right)\right)\right) \,,
 \ee 
are also invariants in this sense. 
 
\label{nonminimal_coupling_definition}
The quantity $\mathcal{I}_1$ can be used to define the notion of nonminimal coupling in an invariant way.
If $\mathcal{I}_1$ is a constant, then 
the scalar field is minimally coupled.
For example, quintessence in general relativity has $\mathcal{A}=1$, $\alpha=0$,
thus $\mathcal{I}_1 \equiv 1$ which holds in any frame and parametrization, i.e. in ``veiled" \cite{deruelle_sasaki} or ``Weyled" \cite{romero} general relativity.
We say that the scalar field is nonminimally coupled if $\mathcal{I}_1' \not\equiv 0$. Later at Eq.~(\ref{scalar_field_equation_without_R}) it becomes clear, that a nonminimally coupled scalar field is sourced by the matter energy-momentum in any frame. If $\mathcal{I}_2\equiv 0$ then the scalar field has a vanishing potential, a property that is not affected by a conformal transformation or reparametrization. Invariant $\mathcal{I}_3$ is given as an indefinite integral and therefore it is constant only if integrand is identically zero. From Eq.~(\ref{semiinvariant_4}) we see that this could only happen if the theory has minimal coupling to curvature and no kinetic term for scalar field. So, in a generic scalar-tensor theory the invariants (\ref{I_1})-(\ref{I_3}) are dynamical functions of $\Phi$, independent of each other.

The assumptions on $\mathcal{A}$ and $\alpha$, listed in Subsec. \ref{general action functional}, bring along a constraint $0<\mathcal{I}_1<\infty$. 
Another useful assumption for the ensuing presentation is 
to demand that $\mathcal{I}_1'$ and $\mathcal{I}_1''$ do not diverge.
In a similar vein also $0\leq\mathcal{I}_2<\infty$, and it makes sense to assume further that $|\mathcal{I}_2^\prime| < \infty$. Constraints on the derivatives are not invariant themselves and therefore we are actually restricting the possible forms of these functions. 

We can also introduce an additional invariant object 
\be
\label{Invariant metric}
\hat{g}_{\mu\nu} \equiv \mathcal{A}(\Phi)g_{\mu\nu} \,,
\ee
which can be used to express geometrical quantities via invariants.
In principle $\hat{g}_{\mu\nu}$ can be considered to be  a metric tensor. 
Note that the choice $\mathcal{A}(\Phi)g_{\mu\nu}$ is not unique. Namely we could have multiplied the metric tensor with any other function of the scalar field which has a suitable transformation property, e.g. 
$e^{2\alpha}g_{\mu\nu}$ \cite{Flanagan}.
The assumption that first and second derivative of $\mathcal{A}$ do not diverge, guarantees that we do not introduce geometrical singularities by defining the invariant metric (\ref{Invariant metric}).

The fact that any function of the invariants is also an invariant can be used to construct further invariants. For example we may define
 \be
 \mathcal{I}_4 \equiv \frac{\mathcal{I}_2}{\mathcal{I}_1^2} = \frac{\mathcal{V}}{e^{4\alpha}}\,.
 \ee 
The transformation of a derivative of these quantities with respect to the scalar field is given by a chain rule,
 \be
 \label{applying chain rule}
 \bar{\mathcal{I}}_i^\prime \equiv \frac{d\bar{\mathcal{I}}_i(\bar{\Phi})}{d\bar{\Phi}} = \frac{d\mathcal{I}_i(\Phi)}{d\Phi}\frac{d\Phi}{d\bar{\Phi}} \equiv \mathcal{I}_i^\prime\bar{f}^\prime \,.
 \ee
 This result is consistent with the transformation properties of a differential of the scalar field,
 \be
 \label{transformation of differential}
 d\bar{\Phi} = \frac{d\bar{\Phi}}{d\Phi}d\Phi \equiv \left( \bar{f}^\prime \right)^{-1}d\Phi \,,
 \ee
 in a sense that integration should cancel differentiation. Indeed
  \be
 \label{integrating derivative of an invariant}
 \bar{\mathcal{I}}_i(\bar{\Phi}) = \mathop{\mathlarger{\int}}\bar{\mathcal{I}}_i^\prime d\bar{\Phi} = \mathop{\mathlarger{\int}} \mathcal{I}_i^\prime \bar{f}^\prime \left( \bar{f}^\prime \right)^{-1}d\Phi = \mathop{\mathlarger{\int}} \mathcal{I}_i^\prime d\Phi = \mathcal{I}_i (\Phi) \,.
 \ee
 Note that we have already used that logic to construct the invariant $\mathcal{I}_3$.
 From Eq. (\ref{applying chain rule}) we conclude that a quotient of the derivatives of invariants is also an invariant
  \be
 \label{quotient of derivatives}
 \mathcal{I}_k = \frac{\mathcal{I}_i^\prime}{\mathcal{I}_j^\prime}\,,
 \ee 
while obviously
 \be
 \label{more_invariants_by_integration}
 \mathcal{I}_i = \mathop{\mathlarger{\int}}\mathcal{I}_k\mathcal{I}_j^\prime d\Phi \,.
 \ee
The expressions (\ref{I_3}) and (\ref{more_invariants_by_integration}) are given in the sense of an antiderivative meaning that they also contain an integration constant. Therefore only their change with respect to some variable should carry physical information.

 By using the rule (\ref{quotient of derivatives}) and a possibility to form arbitrary functions, let us define 
 \be
 \label{I_5}
 \mathcal{I}_5 \equiv \left(\frac{\mathcal{I}_1^\prime}{2\mathcal{I}_1\mathcal{I}_3^\prime}\right)^2 = \frac{\left(2\alpha^\prime\mathcal{A} - \mathcal{A}^\prime\right)^2}{2\mathcal{A}\mathcal{B} + 3\left(\mathcal{A}^\prime\right)^2} \,.
\ee
This invariant helps to distinguish between different theories described by action functional (\ref{fl_moju}). For instance, for minimally coupled scalar field $\mathcal{I}_5 \equiv 0$. For the O'Hanlon type action functional $(\mathcal{B}=0\,,\alpha=0)$ \cite{ohanlon}, which corresponds to the $f(R)$ gravity \cite{dark_energy}, an easy calculation shows that $\mathcal{I}_5 \equiv \frac{1}{3}$. The JF BEPS parametrization is applicable in the range $0 \leq \mathcal{I}_5 < \frac{1}{3}$, while JF BDBW and EF canonical parametrizations cover $0 \leq \mathcal{I}_5 < \infty$. It has been noted before, that in order to match the BDBW parameter range of $-\frac{3}{2}<\omega<0$, the BEPS parametrization should have the sign of the kinetic term flipped \cite{BEPS}.
Violation of the ``no ghosts'' assumption (\ref{no_ghosts}),
corresponding to BDBW $\omega < -\frac{3}{2}$, 
 renders $\mathcal{I}_3$ imaginary and $\mathcal{I}_5$
negative.

In the calculations we sometimes encounter another invariant
 \be
 \label{I_6}
 \mathcal{I}_6 \equiv \left( \frac{\mathcal{I}_2^\prime}{2\mathcal{I}_3^\prime} \right)^2 = \frac{\left(\mathcal{V}^\prime\mathcal{A} - 2\mathcal{V}\mathcal{A}^\prime\right)^2}{\mathcal{A}^4\left(2\mathcal{A}\mathcal{B} + 3\left(\mathcal{A}^\prime\right)^2\right)} \,.
 \ee
The invariants are conveniently summarized in Table \ref{Tabel_1}.

 \subsection{Invariant differential operators}
 Knowledge about the transformation properties of the  differential (\ref{transformation of differential}) allows us to write out invariant differential operators for taking derivatives with respect to the scalar field. These will be in the following form
 \be
 \label{invariant_differentiation_operator}
 \frac{1}{\bar{\mathcal{I}}_j^\prime}\frac{d\phantom{\bar{\Phi}}}{d\bar{\Phi}} = \frac{1}{\mathcal{I}_j^\prime} \frac{d\phantom{\Phi}}{d\Phi} \,.
 \ee
 If we apply that operator to an invariant then the result is also an invariant. For example
 \be
 \frac{1}{\mathcal{I}_j^\prime} \frac{d\phantom{\Phi}}{d\Phi}\mathcal{I}_i = \frac{\mathcal{I}_i^\prime}{\mathcal{I}_j^\prime} \,
 \ee
 and we have once again obtained equation (\ref{quotient of derivatives}). Note that as these invariants have the same argument this result could also be written as a derivative of one invariant with respect to another,
 \be
 \label{Derivative_wrt_invariant}
 \frac{\mathcal{I}_i^\prime}{\mathcal{I}_j^\prime} \equiv \frac{d\Phi}{d\mathcal{I}_j}\frac{d\mathcal{I}_i}{d\Phi} = \frac{d\mathcal{I}_i}{d\mathcal{I}_j} \,.
 \ee
 
 Previous knowledge becomes handy when we want to ``translate'' the results from a distinct parametrization into a general one. 
This procedure is based on fact that in common parametrizations any quantity or operator can be replaced by an invariant which in this parametrization functionally coincides with that quantity or operator.
Namely, if for a fixed parametrization there is an invariant which is a fixed function, then we can construct an invariant differentiation operator (\ref{invariant_differentiation_operator}) which in this parametrization functionally coincides with derivative with respect to scalar field. For example let us take a look at JF BDBW parametrization. We have $\mathcal{I}_1 = \frac{1}{\Psi}$ from where $\Psi = \frac{1}{\mathcal{I}_1}$. Therefore
 \be
 \frac{d\phantom{\Psi}}{d\Psi} = \frac{d\Phi}{d\Psi}\frac{d\phantom{\Phi}}{d\Phi} = \frac{1}{\frac{d\Psi}{d\Phi}}\frac{d\phantom{\Phi}}{d\Phi} = \frac{1}{\frac{d\phantom{\Phi}}{d\Phi}\left(\frac{1}{\mathcal{I}_1}\right)} \frac{d\phantom{\Phi}}{d\Phi} \,.
 \ee
Although that last equality holds only in this parametrization, it allows us to define invariant differentiation operator
 \be
 \label{D_1}
 \mathcal{D}_1 \equiv
 \frac{1}{\frac{d\phantom{\Phi}}{d\Phi}\left(\frac{1}{\mathcal{I}_1}\right)} \frac{d\phantom{\Phi}}{d\Phi} = 
  -\frac{\mathcal{I}_1^2}{\mathcal{I}_1^\prime}\frac{d\phantom{\Phi}}{d\Phi} = \frac{e^{2\alpha}}{\mathcal{A}^\prime - 2\mathcal{A}\alpha^\prime}\frac{d\phantom{\Phi}}{d\Phi} \,,
 \ee
 which in JF BDBW coincides with $\frac{d\phantom{\Psi}}{d\Psi}$.
 Analogically in JF BEPS $\frac{d\phantom{\phi}}{d\phi}$ coincides with
 \be
 \label{D_2}
 \mathcal{D}_2 \equiv \frac{\sqrt{\mathcal{I}_1}}{\mathcal{I}_3^\prime\sqrt{2\left(1-3\mathcal{I}_ 5\right)}}\frac{d\phantom{\Phi}}{d\Phi}  = \pm \frac{e^{\alpha}}{\sqrt{\mathcal{B} - 6 \left(\alpha^\prime\right)^2\mathcal{A} + 6 \alpha^\prime\mathcal{A}^\prime}}\frac{d\phantom{\Phi}}{d\Phi} \,
 \ee
 and in EF canonical parametrization $\frac{d\phantom{\varphi}}{d\varphi}$ coincides with
 \be
 \label{D_3}
 \mathcal{D}_3 \equiv \frac{1}{\mathcal{I}_3^\prime}\frac{d\phantom{\Phi}}{d\Phi} = \pm\frac{2\mathcal{A}}{\sqrt{2\mathcal{A}\mathcal{B} + 3 \left(\mathcal{A}^\prime\right)^2}}\frac{d\phantom{\Phi}}{d\Phi} \,.
 \ee
These results are also gathered in Table
~\ref{Tabel_1}.
Eq.~(\ref{D_2}) tells once again  that JF BEPS is narrower than the other two. The term under the square root must be non-negative and therefore $\mathcal{I}_5<\frac{1}{3}$.

 \renewcommand{\arraystretch}{10}
 \begin{table}
 	\fontsize{9.5}{4}\selectfont
 	\begin{tabular}{|c|c|c|c|c|}
 		\hline
 		{\normalsize Invariant} & {\normalsize General parametrization} & {\normalsize JF BDBW} & {\normalsize JF BEPS} & {\normalsize EF can.} \\
 		\hline
 		$\mathcal{I}_1$ & $\frac{e^{2\alpha(\Phi)}}{\mathcal{A}(\Phi)}$ & $\frac{1}{\Psi}$ & $\frac{1}{F(\phi)}$ & $e^{2\alpha(\varphi)}$ \\
 		\hline
 		$\mathcal{I}_2$ & $\frac{\mathcal{V}(\Phi)}{\mathcal{A}(\Phi)^2}$ & $\frac{\mathcal{V}(\Psi)}{\Psi^2}$ & $\frac{\mathcal{V}(\phi)}{F(\phi)^2}$ & $\mathcal{V}(\varphi)$ \\
 		\hline
 		$\mathcal{I}_3$ & $\pm\mathop{\mathlarger{\int}}\sqrt{\frac{2\mathcal{A}(\Phi)\mathcal{B}(\Phi) + 3 \left(\mathcal{A}^\prime(\Phi)\right)^2}{4\mathcal{A}(\Phi)^2}}d\Phi$ & $\pm\mathop{\mathlarger{\int}}\sqrt{\frac{2\omega(\Psi) + 3}{4\Psi^2}}d\Psi$ & $\pm\mathop{\mathlarger{\int}}\sqrt{\frac{2F(\phi) + 3 \left(F^\prime(\phi)\right)^2}{4F(\phi)^2}}d\phi$ & $\pm\varphi + \mathit{const}$ \\
 		\hline
 		$\mathcal{I}_4\equiv\frac{\mathcal{I}_2}{\mathcal{I}_1^2}$ & $\frac{\mathcal{V}(\Phi)}{e^{4\alpha(\Phi)}}$ & $\mathcal{V}(\Psi)$ & $\mathcal{V}(\phi)$ & $\frac{\mathcal{V}(\varphi)}{e^{4\alpha(\varphi)}}$ \\
 		\hline
 		$\mathcal{I}_5\equiv \left(\frac{\mathcal{I}_1^\prime}{2\mathcal{I}_1\mathcal{I}_3^\prime}\right)^2$ & $\frac{\left(2\alpha^\prime(\Phi)\mathcal{A}(\Phi) - \mathcal{A}^\prime(\Phi)\right)^2}{2\mathcal{A}(\Phi)\mathcal{B}(\Phi) + 3\left(\mathcal{A}^\prime(\Phi)\right)^2}$ & $\frac{1}{2\omega(\Psi) + 3}$ & $\frac{\left(F^\prime(\phi)\right)^2}{2F(\phi) + 3\left(F^\prime(\phi)\right)^2}$ & $\left(\alpha^\prime(\varphi)\right)^2$ \\
 		\hline
 		$\mathcal{I}_6 \equiv \left( \frac{\mathcal{I}_2^\prime}{2\mathcal{I}_3^\prime} \right)^2$ & $\frac{\left(\mathcal{V}^\prime(\Phi)\mathcal{A}(\Phi) - 2\mathcal{V}(\Phi)\mathcal{A}^\prime(\Phi)\right)^2}{\mathcal{A}(\Phi)^4\left(2\mathcal{A}(\Phi)\mathcal{B}(\Phi) + 3\left(\mathcal{A}^\prime(\Phi)\right)^2\right)}$ & $\frac{\left(\mathcal{V}^\prime(\Psi)\Psi -2\mathcal{V}(\Psi)\right)^2}{\Psi^4\left(2\omega(\Psi) + 3\right)}$ & $\frac{\left(\mathcal{V}^\prime(\phi)F(\phi) - 2\mathcal{V}(\phi)F^\prime(\phi)\right)^2}{F(\phi)^4\left(2F(\phi) + 3\left(F^\prime(\phi)\right)^2\right)}$ & $\frac{\left(\mathcal{V}^\prime(\varphi)\right)^2}{4}$ \\
 		\hline
 		$\mathcal{D}_1$ & $\frac{e^{2\alpha(\Phi)}}{\mathcal{A}^\prime(\Phi) - 2\mathcal{A}(\Phi)\alpha^\prime(\Phi)}\frac{d\phantom{\Phi}}{d\Phi}$ & $\frac{d\phantom{\Psi}}{d\Psi}$ & $\frac{1}{F^\prime(\phi)}\frac{d\phantom{\phi}}{d\phi}$ & $-\frac{e^{2\alpha(\varphi)}}{2\alpha^\prime(\varphi)}\frac{d\phantom{\varphi}}{d\varphi}$ \\
 		\hline
 		$\mathcal{D}_2$ & $\frac{\pm e^{\alpha(\Phi)}}{\sqrt{\mathcal{B}(\Phi) - 6 \left(\alpha^\prime(\Phi)\right)^2\mathcal{A}(\Phi) + 6 \alpha^\prime(\Phi)\mathcal{A}^\prime(\Phi)}}\frac{d\phantom{\Phi}}{d\Phi}$ & $\pm \sqrt{\frac{\Psi}{\omega(\Psi)}}\frac{d\phantom{\Psi}}{d\Psi}$ & $\pm\frac{d\phantom{\phi}}{d\phi}$ & $\frac{\pm 2^{-\frac{1}{2}}e^{\alpha(\varphi)}}{\sqrt{1 - 3\left(\alpha^\prime(\varphi)\right)^2}}\frac{d\phantom{\varphi}}{d\varphi}$ \\
 		\hline
 		$\mathcal{D}_3$ & $\pm\frac{2\mathcal{A}(\Phi)}{\sqrt{2\mathcal{A}(\Phi)\mathcal{B}(\Phi) + 3 \left(\mathcal{A}^\prime(\Phi)\right)^2}}\frac{d\phantom{\Phi}}{d\Phi}$ & $\pm\frac{2\Psi}{\sqrt{2\omega(\Psi) + 3}}\frac{d\phantom{\Psi}}{d\Psi}$ & $\pm\frac{2F(\phi)}{\sqrt{2F(\phi) + 3\left(F^\prime(\phi)\right)^2}}\frac{d\phantom{\phi}}{d\phi}$ & $\pm\frac{d\phantom{\varphi}}{d\varphi}$ \\
 		\hline
 		$e^{2\alpha}g_{\mu\nu}$ & $e^{2\alpha(\Phi)}g_{\mu\nu}$ & $g_{\mu\nu}$ & $g_{\mu\nu}$ & $e^{2\alpha(\varphi)}g_{\mu\nu}$ \\
 		\hline 
 		$\mathcal{A}g_{\mu\nu}$ & $\mathcal{A}(\Phi)g_{\mu\nu}$ & $\Psi g_{\mu\nu}$ & $F(\phi)g_{\mu\nu}$ & $g_{\mu\nu}$ \\
 		\hline
 	\end{tabular}
 	\vspace{0.0cm}
 	\caption{Invariants in different parametrizations.}
  	\label{Tabel_1}
 \end{table}
  \renewcommand{\arraystretch}{1}

 \subsection{Invariants in different parametrizations}
 
 The invariants and their functional forms in three  common parametrizations are presented in Table \ref{Tabel_1} which can be used to obtain transformation rules between different parametrizations and the most general one.
 For example if one wants to find a relation between the JF BDBW scalar field $\Psi$ and the EF canonical scalar field $\varphi$ in terms of the JF BDBW variables, then one has to search for an invariant counterpart of the derivative with respect to the EF canonical scalar field $\frac{d\phantom{\varphi}}{d\varphi}$. From that row in Table \ref{Tabel_1} one can write out
 \be
 \pm\frac{2\Psi}{\sqrt{2\omega(\Psi) + 3}}\frac{d\phantom{\Psi}}{d\Psi} = \pm\frac{d\phantom{\varphi}}{d\varphi} \,,
 \ee
hence
  \be
 \label{JF EF can 1}
 \left(\frac{d\Psi}{d\varphi}\right)^2 = \frac{4\Psi^2}{2\omega(\Psi) + 3} \,,
 \ee 
which can be integrated to obtain $\varphi(\Psi)$.
 If we want the same in terms of the EF variables, we should look for an invariant counterpart of the derivative with respect to the JF BDBW scalar field,
 \be
 \frac{d\phantom{\Psi}}{d\Psi} =-\frac{e^{2\alpha(\varphi)}}{2\alpha^\prime(\varphi)}\frac{d\phantom{\varphi}}{d\varphi} \,,
 \ee 
from where 
 \be
 \label{JF EF can 2}
 \left(\frac{d\Psi}{d\varphi}\right)^2 = e^{-4\alpha(\varphi)}4\left(\alpha^\prime(\varphi)\right)^2 \,,
 \ee
which integrates to  $\Psi(\varphi)$.
The relation between the right hand sides of Eqs. (\ref{JF EF can 1}) and (\ref{JF EF can 2}) can be also acquired by combining the rows for $\mathcal{I}_1$ and $\mathcal{I}_5$.
 
Table \ref{Tabel_1} can also be used for transforming invariant quantities from a distinct parametrization to the general one.  
 For example, one may want to quickly find out how the expression $\frac{\left(\mathcal{V}^\prime\Psi -2\mathcal{V}\right)^2}{\Psi^4\left(2\omega + 3\right)}$ written in JF BDBW reads in the general parametrization. Here one should take $\mathcal{I}_4$ which in JF BDBW has the same functional form as potential $\mathcal{V}$, and then apply invariant differentiation $\mathcal{D}_1$ on $\mathcal{I}_4$ to get the invariant counterpart for $\mathcal{V}^\prime$. Further, the invariant $\mathcal{I}_1$ is in this parametrization identical to $\frac{1}{\Psi}$, while $\frac{1}{2\omega + 3}$ should be replaced by $\mathcal{I}_5$. So combining these pieces together gives the whole expression 
 \be
 \left(\mathcal{I}_1^{-1}\mathcal{D}_1\mathcal{I}_4 -2\mathcal{I}_4\right)^2\mathcal{I}_1^4\mathcal{I}_5 = \mathcal{I}_6 \,,
 \ee
where some manipulation and definitions of the invariants have been used on the LHS.
 However, if the quantity we want to transform is not invariant, some caution is needed, since undetermined multiplicative factors of the transformation functions $\bar{f}$ and $\bar{\gamma}$ can be missed out in the procedure.

\subsection{Scalar field $\Phi$ as function of $\mathcal{I}_3$}

In each parametrization we can in principle express $\Phi$ as a function of any invariant $\mathcal{I}_i$. 
Considering the scalar field equation of motion
(\ref{scalar_field_equation_in_terms_of_invariants})
later in the paper, it is useful to express $\Phi$ as a function of $\mathcal{I}_3$. 
In EF canonical parametrization $\mathcal{I}_{3} \sim \varphi$, but
for some other parametrizations (e.g. JF BDBW and JF BEPS) $\mathcal{I}_3$ is given in the form of an indefinite integral (\ref{I_3}) and finding an inverse can be complicated. 
However, under certain conditions we can always approximate $\Phi = \Phi(\mathcal{I}_3)$ as a Taylor expansion around some value $\Phi_0$. 

We start by noticing that $\mathcal{I}_3$ as an indefinite integral contains an integration constant which in principle can be chosen so that $\left.\mathcal{I}_3\right|_{\Phi_0} = 0$. 
Recall that
\be
\frac{d}{d\mathcal{I}_3} = \frac{1}{\mathcal{I}_3^\prime}\frac{d}{d\Phi} \equiv \pm\frac{1}{\sqrt{\mathcal{F}}}\frac{d}{d\Phi} \,.
\ee
Therefore the Taylor expansion reads as follows: 
\beq
\nonumber
\Phi(\mathcal{I}_3) - \Phi_0 & = & \left.\frac{d\Phi}{d\mathcal{I}_3} \right|_{\Phi_0} \cdot \mathcal{I}_3 + \left.\frac{d^2\Phi}{d\mathcal{I}_3^2}\right|_{\Phi_0} \cdot \frac{\mathcal{I}_3^2}{2!} + \ldots = \left.\frac{1}{\mathcal{I}_3^\prime} \right|_{\Phi_0} \cdot \mathcal{I}_3 + \left[ \frac{1}{\mathcal{I}_3^\prime}\frac{d}{d\Phi}\left( \frac{1}{\mathcal{I}_3^\prime} \right) \right]_{\Phi_0} \cdot \frac{\mathcal{I}_3^2}{2!} + \ldots \\
\label{Taylor_series_of_Phi}
&=& \pm\left.\frac{1}{\sqrt{\mathcal{F}}} \right|_{\Phi_0} \cdot \mathcal{I}_3 + \left[ \frac{1}{2}\left(\frac{1}{\mathcal{F}} \right)^\prime \right]_{\Phi_0} \cdot \frac{\mathcal{I}_3^2}{2!} + \ldots 
\eeq
where we have used 
\be
\label{derivative_of_one_over_sqrt}
\frac{1}{\sqrt{\mathcal{F}}}\frac{d}{d\Phi} \left( \frac{1}{\sqrt{\mathcal{F}}} \right) \equiv \frac{1}{\sqrt{\mathcal{F}}}\left( \frac{1}{\sqrt{\mathcal{F}}} \right)^\prime = -\frac{1}{2}\frac{\mathcal{F}^\prime}{\mathcal{F}^2} = \frac{1}{2} \left( \frac{1}{\mathcal{F}} \right)^\prime \,.
\ee 
One can show that the coefficients in the Taylor series (\ref{Taylor_series_of_Phi}) do not diverge and at least some of them are non-vanishing if
\beq
\label{nonvanishing_F}
0 \leq \frac{1}{\mathcal{F}} &<& \infty \,,\\
\label{non_diverging_derivatives_of_1_over_F}
-\infty < \left(\frac{1}{\mathcal{F}}\right)^{\overset{n \text{-times}}{\overbrace{\prime \dots \prime}}}  &\equiv&  \frac{d^n}{d\Phi^n}\left(\frac{1}{\mathcal{F}}\right) < \infty \,,\\
\label{non_vanishing_first_derivative_of_1_over_F}
\text{if } \left.\frac{1}{\mathcal{F}}\right|_{\Phi_0} &=& 0 \,, \text{ then } \left.\left(\frac{1}{\mathcal{F}}\right)^\prime\right|_{\Phi_0} \neq 0 \,.
\eeq
The same assumptions arose in context of Friedmann cosmology \cite{JKS2008}.
They restrict the possible forms of $\mathcal{F} \equiv (\mathcal{I}_3')^2 $
and the scalar field dynamics. 
These assumptions complement the restrictions on $\mathcal{I}_1$, $\mathcal{I}_2$ and their derivatives discussed earlier.
A few comments follow.

First, the assumption (\ref{non_vanishing_first_derivative_of_1_over_F}) imposes that $\left.\frac{1}{\mathcal{F}}\right|_{\Phi_0} = 0$ is not an extremum. Therefore, if the scalar field $\Phi$ would evolve through value $\Phi_0$ then $\mathcal{F}$ would go negative, thereby violating the condition (\ref{nonvanishing_F}), i.e.\ (\ref{no_ghosts}). 
A consistent theory would avoid this to happen. 
Indeed, if the linear term in the Taylor expansion (\ref{Taylor_series_of_Phi}) vanishes due to $\frac{1}{\mathcal{F}}=0$, then the assumption (\ref{non_vanishing_first_derivative_of_1_over_F})
guarantees that the coefficient of the quadratic term is definitely non-vanishing,
\be
\label{non_crossing_condition}
\Phi(\mathcal{I}_3)-\Phi_0 \approx \left.\frac{1}{4}\left(\frac{1}{\mathcal{F}}\right)^\prime\right|_{\Phi_0} \mathcal{I}_3^2.
\ee
Hence the possible scalar field $\Phi$ values are never smaller (higher) than $\Phi_0$ if $\left.\left( \frac{1}{\mathcal{F}}\right)^\prime \right|_{\Phi_0}$ is positive (negative), which means that the scalar field $\Phi$ can approach $\Phi_0$ from above (form below).

Second, since here $\mathcal{I}_3$ is an invariant infinitesimal quantity, we can use it as a scale to compare the order of magnitude of the perturbation $\Phi(\mathcal{I}_3)-\Phi_0$ in different parametrizations. In the parametrization where $\left.\mathcal{F}\right|_{\Phi_0}$ is regular 
the Taylor expansion (\ref{Taylor_series_of_Phi}) starts with a linear term and the perturbation $\Phi(\mathcal{I}_3)-\Phi_0$ is the same order small as $\mathcal{I}_3$. While expanding at $\bar{\Phi}_0= f(\Phi_0)$ in another parametrization,  if  $\left.\bar{\mathcal{F}}\right|_{\bar{\Phi}_0}$ diverges, the corresponding  perturbation $\bar{\Phi}(\mathcal{I}_3)-\bar{\Phi}_0$ is quadratically small compared to $\mathcal{I}_3$, 
as Eq.~(\ref{non_crossing_condition}).

Third, if the leading coefficient in the Taylor series
(\ref{Taylor_series_of_Phi}) vanishes, then $\left.\nabla_\mu \Phi\right|_{\Phi_0}=0$, because $\left.\mathcal{I}_3\right|_{\Phi_0}=0$ and
\be
	\left.\nabla_\mu \Phi\right|_{\Phi_0} \approx \nabla_\mu \left[ \left. \frac{1}{2}\left(\frac{1}{\mathcal{F}} \right)^\prime \right|_{\Phi_0} \cdot \frac{\mathcal{I}_3^2}{2!} \right]_{\Phi_0} = \left[ \left. \frac{1}{2}\left(\frac{1}{\mathcal{F}} \right)^\prime \right|_{\Phi_0} \cdot \mathcal{I}_3 \right]_{\mathcal{I}_3|_{\Phi_0}}\left.\nabla_\mu\mathcal{I}_3 \right|_{\Phi_0} = 0 \,
\ee
even if $\left. \nabla_\mu\mathcal{I}_3 \right|_{\Phi_0} \neq 0$.

Finally, we may remark that since in the Einstein frame canonical parametrization $\mathcal{I}_3 = \pm \varphi + \mathit{const}$ all discussion in this subsection is equivalent to the Taylor expansion of the general scalar field $\Phi$ as a function of the EF canonical scalar field $\varphi$. 

\section{Equations of motion}

\subsection{Equations of motion in the general parametrization}
Varying the action (\ref{fl_moju}) with respect to the metric tensor gives a tensor equation
\beq
\nonumber
&&\mathcal{A}\left(R_{\mu\nu} - \frac{1}{2}g_{\mu\nu}R\right) + \left(\frac{1}{2}\mathcal{B}+\mathcal{A}^{\prime\prime}\right) g_{\mu\nu} g^{\rho\sigma}\nabla_\rho\Phi\nabla_\sigma\Phi - \left(\mathcal{B} + \mathcal{A}^{\prime\prime}\right)\nabla_\mu\Phi\nabla_\nu\Phi  \\
\label{tensor_equation}
&&\hspace{2cm}+ \mathcal{A}^\prime\left(g_{\mu\nu}\Box\Phi - \nabla_\mu\nabla_\nu\Phi\right) + \frac{1}{\ell^2} g_{\mu\nu}\mathcal{V} - \kappa^2T_{\mu\nu} = 0\,,
\eeq
where the matter energy-momentum tensor is
\be
T_{\mu\nu} \equiv -\frac{2}{\sqrt{-g}}\frac{\delta S_m}{\delta g^{\mu\nu}} \,.
\ee
Analogously, varying the action (\ref{fl_moju}) with respect to the scalar field gives us an equation of motion for the scalar field
\be
\label{scalar field equation containing R}
R\mathcal{A}^\prime + \mathcal{B}^\prime g^{\mu\nu}\nabla_\mu\Phi\nabla_\nu\Phi + 2\mathcal{B}\Box\Phi - 2\ell^{-2}\mathcal{V}^\prime + 2\kappa^2\alpha^\prime T = 0  \,,
\ee
where $T\equiv g^{\mu\nu}T_{\mu\nu}$.
Using the trace of the tensor equation (\ref{tensor_equation}) to eliminate $R$ from the scalar field equation (\ref{scalar field equation containing R}) yields
\be
\frac{2\mathcal{A}\mathcal{B} + 3\left( \mathcal{A}^\prime\right)^2}{\mathcal{A}}\Box\Phi + \frac{\left( 2\mathcal{A}\mathcal{B} + 3\left( \mathcal{A}^\prime \right)^2 \right)^\prime}{2\mathcal{A}} g^{\mu\nu} \nabla_\mu\Phi \nabla_\nu\Phi - \frac{2\left( \mathcal{A}\mathcal{V}^\prime - 2\mathcal{A}^\prime \mathcal{V} \right)}{\ell^2\mathcal{A}} + \frac{\kappa^2\left( 2\mathcal{A}\alpha^\prime - \mathcal{A}^\prime \right)}{\mathcal{A}}T = 0 \,. 
\label{scalar_field_equation_without_R}
\ee
As alluded before in Subsec. \ref{nonminimal_coupling_definition},
one way to define the meaning of nonminimal coupling is that the scalar field in Eq.~(\ref{scalar_field_equation_without_R}) is sourced by 
the contracted matter energy-momentum tensor $T$. 
Inspection of the last term on LHS confirms the claim that nonminimal coupling is realized when  $\mathcal{I}_1' \not\equiv 0$. 
The continuity equation  
\be
\label{EI_jaavus}
\nabla^\mu T_{\mu\nu} = \alpha^\prime T \nabla_\nu\Phi \,
\ee
tells that the usual matter energy-momentum is covariantly conserved in those parametrizations where $\alpha(\Phi)=const$.

\subsection{Equations of motion in terms of the invariants}

We have noted that the invariant object $\hat{g}_{\mu\nu} \equiv \mathcal{A} g_{\mu\nu}$, introduced in Eq.~(\ref{Invariant metric}), 
can be taken as a metric tensor and therefore it is possible to calculate Christoffel symbols with respect to it,
\be
\label{Invariant_Christoffel_symbols}
\hat{\Gamma}^{\lambda}_{\mu\nu} = \Gamma^{\lambda}_{\mu\nu} +\frac{\mathcal{A}^\prime}{2\mathcal{A}}\left( \delta^\lambda_\mu \partial_\nu\Phi + \delta^\lambda_\nu \partial_\mu\Phi - g_{\mu\nu}g^{\lambda\sigma}\partial_\sigma\Phi \right) \,.
\ee
Mathematically this result is the well known transformation rule for Christoffel symbols under the conformal transformation \cite{stg_books,Wald},
or the definition corresponding to Weyl-integrable geometry \cite{quiros, romero}. 
But here the point is simply that $\hat{\Gamma}^{\lambda}_{\mu\nu}$ 
remains invariant under the transformations (\ref{conformal_transformation}) and (\ref{field_redefinition}). Now we can use (\ref{Invariant_Christoffel_symbols}) to define covariant derivative with respect to $\hat{g}_{\mu\nu}$, e.g. $\hat{\nabla}_\mu V^\nu = \partial_\mu V^\nu + \hat{\Gamma}^\nu_{\mu\lambda}V^\lambda$ etc.
Similarly
the objects $\hat{\Gamma}^{\lambda}_{\mu\nu}$ can be employed to build the Riemann-Christoffel tensor $\hat{R}^{\lambda}_{\,\,\, \mu \rho \nu}$ which in this case is manifestly invariant under conformal transformation and scalar field reparametrization. Therefore we can also construct the Einstein tensor $\hat{G}_{\mu\nu} \equiv \hat{R}_{\mu\nu} - \frac{1}{2} \hat{g}_{\mu\nu} \hat{R}$ which can be expressed in terms of $g_{\mu\nu}$ and $\mathcal{A}(\Phi)$ as 
\beq
\nonumber
\hat{G}_{\mu\nu} &=& R_{\mu\nu}-\frac{1}{2}g_{\mu\nu}R + \frac{\mathcal{A}^{\prime\prime}}{\mathcal{A}}g_{\mu\nu} g^{\rho\sigma}\nabla_\rho\Phi \nabla_\sigma\Phi - \frac{\mathcal{A}^{\prime\prime}}{\mathcal{A}} \nabla_\mu\Phi \nabla_\nu\Phi + \frac{\mathcal{A}^\prime}{\mathcal{A}}g_{\mu\nu}\Box\Phi - \frac{\mathcal{A}^\prime}{\mathcal{A}}\nabla_\mu \nabla_\nu\Phi  \\ 
\label{Invariant_Riemann_Christoffel_tensor}
&&\hspace{0.5cm}-\frac{3\left( \mathcal{A}^\prime \right)^2}{4\mathcal{A}^2}g_{\mu\nu}g^{\rho\sigma}\nabla_\rho \Phi \nabla_\sigma\Phi + \frac{3\left( \mathcal{A}^\prime \right)^2}{2\mathcal{A}^2} \nabla_\mu\Phi \nabla_\nu\Phi\,.
\eeq  
In the same spirit we can define an energy-momentum tensor that is invariant under conformal transformation and scalar field reparametrization
\be
\label{Invariant_EI_tensor}
\hat{T}_{\mu\nu} \equiv -\frac{2}{\sqrt{-\hat{g}}}\frac{\delta S_m}{\delta \hat{g}^{\mu\nu}} = -\frac{2}{\mathcal{A}^2\sqrt{-g}} \frac{\delta g^{\mu\nu}}{\delta \hat{g}^{\mu\nu}} \frac{\delta S_m}{\delta g^{\mu\nu}} = \frac{1}{\mathcal{A}}\left\lbrace -\frac{2}{\sqrt{-g}}\frac{\delta S_m}{\delta g^{\mu\nu}} \right\rbrace \equiv \frac{1}{\mathcal{A}}T_{\mu\nu} \,.
\ee

Comparing the result (\ref{Invariant_Riemann_Christoffel_tensor}) with Eq.~(\ref{tensor_equation}) while taking into account the definitions (\ref{semiinvariant_4}), (\ref{I_2}), (\ref{I_3}) and (\ref{Invariant_EI_tensor}) allows us to rewrite Eq. (\ref{tensor_equation}) as follows
\be
\label{tensor_equation_in_terms_of_invariants}
\mathcal{A}\left\lbrace \hat{G}_{\mu\nu} + \hat{g}_{\mu\nu} \hat{g}^{\rho\sigma} \hat{\nabla}_\rho \mathcal{I}_3 \hat{\nabla}_\sigma \mathcal{I}_3 - 2\hat{\nabla}_\mu\mathcal{I}_3 \hat{\nabla}_\nu\mathcal{I}_3 + \ell^{-2}\hat{g}_{\mu\nu}\mathcal{I}_2 - \kappa^2 \hat{T}_{\mu\nu} \right\rbrace = 0 \,,
\ee
and the scalar field equation  (\ref{scalar_field_equation_without_R}) as
\be
\label{scalar_field_equation_without_R_with_invariants}
4\mathcal{A}^2 \left\lbrace \mathcal{F} \hat{\rule{0ex}{1.5ex}\Box}\Phi + \frac{1}{2}\mathcal{F}^\prime\hat{g}^{\mu\nu} \hat{\nabla}_\mu\Phi \hat{\nabla}_\nu\Phi - \frac{1}{2\ell^2} \mathcal{I}_2^\prime + \frac{\kappa^2}{4}  \frac{ \mathcal{I}_1^\prime}{\mathcal{I}_1}\hat{T} \right\rbrace =0 \,.
\ee
Here $\Box$ operator with respect to $\hat{g}_{\mu\nu}$ is defined by 
\beq
\hat{\rule{0ex}{1.5ex}\Box} \Phi &\equiv& 
\frac{1}{\sqrt{-\hat{g}}}\partial_\mu\left( \sqrt{-\hat{g}} \hat{g}^{\mu\nu} \partial_\nu\Phi \right) 
= \frac{\mathcal{A}^\prime}{\mathcal{A}^2}g^{\mu\nu} \nabla_\mu\Phi \nabla_\nu\Phi + \frac{1}{\mathcal{A}}\Box\Phi \,.
\eeq
Due to the identity
\be
\mathcal{I}_3^\prime \hat{\rule{0ex}{1.5ex}\Box}\mathcal{I}_3 = \mathcal{F}\hat{\rule{0ex}{1.5ex}\Box}\Phi + \frac{1}{2}\mathcal{F}^\prime \hat{g}^{\mu\nu}\hat\nabla_\mu\Phi \hat\nabla_\nu\Phi \,
\ee
we may write the scalar field equation (\ref{scalar_field_equation_without_R_with_invariants})
as
\be
\label{scalar_field_equation_in_terms_of_invariants}
4\mathcal{A}^2\mathcal{I}_3^\prime \left\lbrace \hat{\rule{0ex}{1.5ex}\Box}\mathcal{I}_3 -  \frac{1}{2\ell^2} \frac{\mathcal{I}_2^\prime}{\mathcal{I}_3^\prime} + \frac{\kappa^2}{4\mathcal{I}_1} \frac{\mathcal{I}_1^\prime}{\mathcal{I}_3^\prime} \hat{T} \right\rbrace =0 \,.
\ee
Since by the assumption neither $\mathcal{A}$ nor $\mathcal{I}_3'= \pm \sqrt{\mathcal{F}}$ can vanish, we can divide the last equation with term in front of the braces, and obtain an equation where each term is an invariant,
\be
\label{Invariant_Klein_Gordon_equation}
\hat{\rule{0ex}{1.5ex}\Box}\mathcal{I}_3 -  \frac{1}{2\ell^2} \frac{d\mathcal{I}_2}{d\mathcal{I}_3} + \frac{\kappa^2}{4} \frac{d\ln\mathcal{I}_1}{d\mathcal{I}_3} \hat{T} = 0 \,.
\ee 
The logic of differentiation used here was introduced before equation (\ref{Derivative_wrt_invariant}). 

\subsection{Action in terms of the invariants}
The definition of conformally invariant metric tensor $\hat{g}_{\mu\nu} \equiv \mathcal{A}g_{\mu\nu}$ was based on the knowledge about transformation properties of $\mathcal{A}$ 
given by (\ref{fl_fnide_teisendused}), which were read off from the transformed action functional (\ref{fl_teisendatud_moju}). 
Therefore it is natural that we can also rewrite the action functional in terms of invariants up to a boundary term, namely 
\beq
\nonumber
S &=& \frac{1}{2\kappa^2}\mathop{\mathlarger{\int_{V_4}}}d^4x\sqrt{-\hat{g}}\left\lbrace \hat{R} - 2\hat{g}^{\mu\nu}\hat{\nabla}_\mu\mathcal{I}_3 \hat{\nabla}_\nu\mathcal{I}_3 - 2\ell^{-2}\mathcal{I}_2 \right\rbrace + S_m\left[ \mathcal{I}_1 \hat{g}_{\mu\nu} , \chi \right]  \\
\label{invariant_action_functional}
&&+ \frac{3}{2\kappa^2}\mathop{\mathlarger{\int_{V_4}}}d^4x \, \partial_\mu \left( \sqrt{-\hat{g}} \hat{g}^{\mu\nu}\partial_\nu \ln \mathcal{A} \right) \,.
\eeq
Varying (\ref{invariant_action_functional}) with respect to $\hat{g}^{\mu\nu}$ and $\mathcal{I}_3$ gives us invariant expressions that coincide with terms in braces in respectively Eq. (\ref{tensor_equation_in_terms_of_invariants}) and (\ref{scalar_field_equation_in_terms_of_invariants}).

As already mentioned, the choice $\hat{g}_{\mu\nu} \equiv \mathcal{A}g_{\mu\nu}$ is not unique, it just seems to give the equations in the simplest form. Note that these expressions remind the Einstein frame equations, because in the Einstein frame the invariant metric $\left.\hat{g}_{\mu\nu}\right|_{EF}$ coincides with the Einstein frame metric $g_{\mu\nu}$, while the invariant $\left.\mathcal{I}_2\right|_{EF}$ coincides with the Einstein frame potential $\mathcal{V}$. If we had chosen $\hat{g}_{\mu\nu} = e^{2\alpha}g_{\mu\nu}$ then equations would have been more similar to the Jordan frame ones.


\section{PPN parameters}

\subsection{PPN parameters in the JF BDBW parametrization}
The aim of this section is to use Table \ref{Tabel_1} for writing the effective gravitational constant $G_{\mathrm{eff}}$ and the parametrized post-Newtonian parameters $\gamma$ and $\beta$ in terms of the invariants and thereby obtain a form which easily allows to get the PPN parameters in any other parametrization. We start from JF BDBW parametrization where the most general calculation was recently accomplished \cite{Hohmann:2013rba}, expanding earlier Refs.~\cite{ppn_bdbw}. Table \ref{Tabel_1} contains all possible objects occurring in that parametrization. We proceed under the premise that PPN parameters are invariants and must be determined uniquely. It does not matter whether we use the transformations (\ref{fl_fnide_teisendused}) to obtain the results in the general parametrization or substitute in the respective invariants 
from Table \ref{Tabel_1}
in order to get an invariant which in a parametrization at hand functionally coincides with PPN parameters.

So, from Ref.~\cite{Hohmann:2013rba} we take following results calculated in the JF BDBW parametrization. The PPN ansatz assumes that in the absence of any perturbation we have flat Minkowski geometry as a background, which leads to the conditions $\mathcal{V}=0$ and $\mathcal{V}^\prime=0$. Taking these conditions into account in the calculation, gives a result which is expressed in terms of the scalar field effective mass
\be
m_{\Psi} \equiv \frac{1}{\ell} \sqrt{\frac{2 \Psi}{2\omega(\Psi) + 3}\frac{d^2 \mathcal{V}}{d\Psi^2} } \,.
\ee
The effective gravitational constant that in an experimental setup multiplies the nonvarying constant $\frac{\kappa^2}{8 \pi}$, and the PPN parameters are given by
\beq
G_{\mathrm{eff}} &=& \frac{1}{\Psi} \left( 1+ \frac{e^{-m_{\Psi}r}}{2\omega + 3} \right) \,,\\
\gamma-1 &=& -\frac{2e^{-m_\Psi r}}{G_{\mathrm{eff}} \Psi (2\omega + 3)} \,, \\
\beta-1 &=& \frac{ \frac{d\omega}{d\Psi}  e^{-2m_{\Psi}r} }{G_{\mathrm{eff}}^2 \Psi (2\omega + 3)^3} - \frac{ m_{\Psi} r}{ G_{\mathrm{eff}}^2 \Psi^2 (2\omega + 3)} \beta(r) \,,
\eeq
where the extra radius dependent contribution in $\beta$, 
\beq
\beta(r) &=&
\frac{1}{2}e^{-2m_{\Psi}r} + \left(m_{\Psi}r + e^{m_{\Psi}r}\right)\mathrm{Ei}(-2m_{\Psi}r) - e^{-m_{\Psi}r}\ln(m_{\Psi}r)  \nonumber \\
&& \,\, + \frac{3\Psi}{2(2\omega + 3)}\left(\frac{\frac{d^3V}{d\Psi^3} }{3 \frac{d^2V}{d\Psi^2}} - \frac{1}{\Psi} - \frac{\frac{d\omega}{d\Psi}}{2\omega + 3}\right)\left(e^{m_{\Psi}r}\mathrm{Ei}(-3m_{\Psi}r) - e^{-m_{\Psi}r}\mathrm{Ei}(-m_{\Psi}r)\right) ,
\eeq
involves exponential integrals $\mathrm{Ei}(m_{\Psi}r)$.
It is understood in these formulas that $\Psi$ and the functions $\omega(\Psi)$, $V(\Psi)$, etc., are all evaluated at the spatially asymptotic constant background value of $\Psi$.

\subsection{PPN parameters in terms of the invariants} 

Let us rewrite the previous result in terms of invariants by making use of Table \ref{Tabel_1}.
The first constraint arising from Minkowskian boundary conditions, $\mathcal{V}=0$, translates into $\mathcal{I}_4 \equiv \frac{\mathcal{I}_2}{\mathcal{I}_1^2} = 0 $ which implies $\mathcal{I}_2 = 0$. 
The second condition $\mathcal{V}^\prime = 0$ gives $\left.\mathcal{D}_1\mathcal{I}_4\right|_{\mathcal{I}_2=0} \equiv \frac{\mathcal{I}_2^\prime}{\mathcal{I}_1^\prime}=0$.
Similarly, the scalar field effective mass reads
\be
m_\Phi = \frac{1}{\ell}\sqrt{2\mathcal{I}_1^{-1}\mathcal{I}_5 \mathcal{D}_1^2\mathcal{I}_4}= \frac{1}{\ell}\sqrt{\frac{\mathcal{I}_2^{\prime\prime}}{2\mathcal{I}_1 \left( \mathcal{I}_3^\prime \right)^2}} \,.
\ee
Here in order to preserve simpler form of the expression at the RHS we have substituted the Minkowskian boundary conditions written in terms of invariants. The quantity on the RHS is invariant only under these conditions.
The effective gravitational constant and the PPN parameters $\gamma$ and $\beta$ turn out to be 
\beq
G_{\mathrm{eff}} &=& \mathcal{I}_1 \left(1 + \mathcal{I}_5 e^{-m_\Phi r} \right) \,,\\
\gamma-1 &=& -\frac{2e^{-m_\Phi r}}{G_{\mathrm{eff}}}\mathcal{I}_1\mathcal{I}_5 \,,\\
\nonumber
\beta - 1 &=& \frac{e^{-2m_\Phi r}}{G_{\mathrm{eff}}^2 \mathcal{I}_1^{-1}}\mathcal{I}_5^3 \left[\mathcal{D}_1\left( \frac{1}{2} \left( \frac{1}{\mathcal{I}_5} - 3 \right) \right)\right] 
- \frac{m_\Phi r}{G_{\mathrm{eff}}^2\mathcal{I}_1^{-2}} \mathcal{I}_5 \, \beta(r)  \\
&=&\frac{1}{2}\frac{\mathcal{I}_1^3\mathcal{I}_5}{G_{\mathrm{eff}}^2} \frac{\mathcal{I}_5^\prime}{\mathcal{I}_1^\prime}  e^{-2 m_\Phi r} -  \frac{m_\Phi r}{G_{\mathrm{eff}}^2} \mathcal{I}_1^{2}\mathcal{I}_5 \, \beta(r) \,,
\eeq
where
\beq
\beta(r) &=&
\frac{1}{2}e^{-2m_{\Phi}r} + \left(m_{\Phi}r + e^{m_{\Phi}r}\right)\mathrm{Ei}(-2m_{\Phi}r) - e^{-m_{\Phi}r}\ln(m_{\Phi}r)  \nonumber \\
&& + \frac{3\mathcal{I}_5}{2\mathcal{I}_1}\left( \frac{1}{3} \frac{\mathcal{D}_1^3\mathcal{I}_4 }{\mathcal{D}_1^2\mathcal{I}_4} - \mathcal{I}_1 - 
\mathcal{I}_5\mathcal{D}_1\left(\frac{1}{2}\left(\frac{1}{\mathcal{I}_5}-3 \right)\right)\right)\left(e^{m_{\Phi}r}\mathrm{Ei}(-3m_{\Phi}r) - e^{-m_{\Phi}r}\mathrm{Ei}(-m_{\Phi}r)\right) 
\,.
\eeq
Note, that the quantity $m_{\Phi} r$ is invariant in our conventions.

\subsection{PPN parameters in the general parametrization}
In the general parametrization, expressing the invariants in terms of the functions $\left\lbrace \mathcal{A},\,\mathcal{B},\, \mathcal{V} ,\, \alpha \right\rbrace$, the result is \cite{JKSV_1}
\be
\label{scalar_field_mass}
m_\Phi = \frac{1}{\ell}\sqrt{ e^{-2\alpha} \frac{2\mathcal{A}\mathcal{V}^{\prime\prime}}{2\mathcal{A}\mathcal{B} + 3\left( \mathcal{A}^\prime\right)^2}} \,
\ee
and
\beq
G_{\mathrm{eff}} &=& \frac{e^{2\alpha}}{\mathcal{A}} \left(1 + \frac{\left(2\alpha^\prime\mathcal{A} - \mathcal{A}^\prime\right)^2}{2\mathcal{A}\mathcal{B} + 3\left(\mathcal{A}^\prime\right)^2} e^{-m_\Phi r} \right) \,,\\
\gamma-1 &=& -\frac{2e^{-m_\Phi r}}{G_{\mathrm{eff}}}\frac{e^{2\alpha}}{\mathcal{A}}\frac{\left(2\alpha^\prime\mathcal{A} - \mathcal{A}^\prime\right)^2}{(2\mathcal{A}\mathcal{B} + 3\left(\mathcal{A}^\prime\right)^2)} \,,\\
\beta - 1 &=& \frac{e^{4\alpha}}{2\mathcal{A} G_{\mathrm{eff}}^2}\frac{\left(2\alpha^\prime\mathcal{A} -\mathcal{A}^\prime\right)}{\left(2\mathcal{A}\mathcal{B} \negthickspace 
\,+\,\negthickspace 3\left(\mathcal{A}^\prime\right)^2\right)} \negthickspace \left(\frac{ \left( 2\alpha^\prime\mathcal{A} -\mathcal{A}^\prime \right)^2}{2\mathcal{A}\mathcal{B} \,\negthickspace + \,\negthickspace 3\left(\mathcal{A}^\prime\right)^2} \right)^\prime e^{-2m_{\Phi}r} \negthickspace - \negthickspace\, \frac{m_\Phi r}{G_{\mathrm{eff}}^2}\frac{e^{4\alpha}}{\mathcal{A}^2} \frac{\left(2\alpha^\prime\mathcal{A} - \mathcal{A}^\prime\right)^2}{\left(2\mathcal{A}\mathcal{B}\,\negthickspace + \,\negthickspace 3\left(\mathcal{A}^\prime\right)^2\right)} \, \beta(r) .
\eeq
Now it is easy to check that there is a match with the Einstein frame calculation \cite{ppn_ef, scharer} and the corresponding expression in the JF BEPS parametrization \cite{BEPS}. The effective mass (\ref{scalar_field_mass}) differs from the one obtained in Ref.~\cite{scharer} by the factor $e^{-\alpha}$, but in the conventions of Ref.~\cite{scharer} this is precisely the factor that relates the masses in Jordan and Einstein frames.


\section{Cosmological solutions}
\subsection{Equations for flat FLRW cosmology without matter}
Let us start with flat ($k=0$) Friedmann-Lema\^itre-Robertson-Walker (FLRW) line element
\be
\label{FLRW line element}
ds^2 \equiv g_{\mu\nu}dx^\mu dx^\nu = -{dt}^2 + \left(a(t)\right)^2\left\lbrace dr^2 + r^2d\vartheta^2 + r^2\sin^2\vartheta d\varphi^2 \right\rbrace \,.
\ee
Now take the conformally invariant metric tensor $\hat{g}_{\mu\nu} \equiv \mathcal{A}g_{\mu\nu}$, Eq.~(\ref{Invariant metric}), where $g_{\mu\nu}$ is in the FLRW form. In order to have $\hat{g}_{\mu\nu}$ also in that form, we should make a coordinate transformation and the scale factor redefinition
\beq
\label{coordinate_transformation_t_to_t_hat}
\frac{d \phantom{\hat{t}}}{d\hat{t}} \equiv \frac{1}{\sqrt{\mathcal{A}}}\frac{d \phantom{{t}}}{dt} \,,\\
\hat{a}(\hat{t} ) \equiv \sqrt{\mathcal{A}} \, a(t) \,.
\eeq
The Hubble parameter $\hat{H}$ calculated in terms of the invariant variables is related to the Hubble parameter $H$ calculated in the frame defined by $g_{\mu\nu}$ as 
\be
\hat{H} \equiv \frac{1}{\sqrt{\mathcal{A}}}\left(H + \frac{1}{2}\frac{\mathcal{A}^\prime}{\mathcal{A}}\dot{\Phi}\right) \,.
\ee
Plugging the invariant form of FLRW metric (\ref{FLRW line element}) into equations (\ref{tensor_equation_in_terms_of_invariants}) and (\ref{Invariant_Klein_Gordon_equation}) yields
\beq
\label{Friedmanns_constraint_in_terms_of_invariants} \hat{H}^2 &=& \frac{1}{3}\left(\frac{d}{d\hat{t}} \mathcal{I}_3 \right)^2 + \frac{1}{3\ell^2}\mathcal{I}_2 \,, \\
\label{Friedmanns_second_equation_in_terms_of_invariants}
2\frac{d}{d\hat{t}} \hat{H} + 3 \hat{H}^2  &=&  -\left(\frac{d}{d\hat{t}}\mathcal{I}_3\right)^2 + \frac{ 1 }{\ell^2}\mathcal{I}_2   \,, \\
\label{Friedmanni skalaarvalja vorrand} 
\frac{d}{d\hat{t}} \left( \frac{d}{d\hat{t}}\mathcal{I}_3 \right)  &=& -3\hat{H} \frac{d}{d\hat{t}}\mathcal{I}_3 - \frac{1}{2\ell^2}\frac{\mathcal{I}_2^\prime}{\mathcal{I}_3^\prime} \,.
\eeq
We have dropped the matter terms, i.e. $\hat{T}_{\mu\nu}\equiv 0$. By doing this we have truncated the theory by omitting $\alpha$, thus we are left with only three arbitrary functions $\left\{ \mathcal{A}\,,\mathcal{B}\,,\mathcal{V} \right\}$.

\subsection{Scalar field equation as a dynamical system}

The first equation of the system (\ref{Friedmanns_constraint_in_terms_of_invariants})-(\ref{Friedmanni skalaarvalja vorrand}) 
is a constraint, therefore we may focus only upon Eq.~(\ref{Friedmanni skalaarvalja vorrand}) where 
the geometrical quantity $\hat{H}$ has been substituted from Eq.~(\ref{Friedmanns_constraint_in_terms_of_invariants}),
\be
\label{skalaarvalja_kosmo_vorrand_invariantidena}
\frac{d}{d\hat{t}} \left( \frac{d}{d\hat{t}}\mathcal{I}_3 \right)  = - \varepsilon \sqrt{3\left(\frac{d}{d\hat{t}} \mathcal{I}_3 \right)^2 + \frac{3}{\ell^{2}}\mathcal{I}_2}\,\,\frac{d}{d\hat{t}}\mathcal{I}_3 - \frac{1}{2\ell^2}\frac{d\mathcal{I}_2}{d\mathcal{I}_3} \,,
\ee
where $\varepsilon=+1$ ($\varepsilon=-1$) corresponds to an expanding (contracting) universe with respect to the metric  $\hat{g}_{\mu\nu}$.
In order to learn about the general features of the cosmological solutions it is instructive to write 
the scalar field equation as a dynamical system
and ask whether there are any fixed points and what are their properties.
For $\Phi_0$ to give a fixed point we must insist that $\left. \frac{d}{d\hat{t}}\mathcal{I}_3 \right|_{\Phi_0} = 0$ and $\left. \frac{d^2}{d\hat{t}^2}\mathcal{I}_3 \right|_{\Phi_0} = 0$ . From Eq. (\ref{skalaarvalja_kosmo_vorrand_invariantidena}) we see
that this occurs when
\be
\label{fixed_point_condition}
\left.\frac{d\mathcal{I}_2}{d\mathcal{I}_3}\right|_{\Phi_0} \equiv
\left. \frac{\mathcal{I}_2^\prime}{\mathcal{I}_3^\prime} \right|_{\Phi_0} = 0 \,.
\ee
Hereby we may distinguish two types of the scalar field  values $\Phi_0$:
\beq
\label{critical_point_moon}
\Phi_\bullet &:& \left.\mathcal{I}_2^\prime\right|_{\Phi_\bullet}=0\,, \qquad \left. \frac{1}{\mathcal{I}_3^\prime} \right|_{\Phi_\bullet}\neq 0 \,, \\
\label{critical_point_star}
\Phi_\star   &:& \left. \frac{1}{\mathcal{I}_3^\prime} \right|_{\Phi_\star}=0 \,.
\eeq 
Note that the condition (\ref{fixed_point_condition}) for a fixed point is invariant, while the distinction (\ref{critical_point_moon}), (\ref{critical_point_star}) is not. 
Therefore, if a fixed point occurs in some parametrization, then a corresponding fixed point will be present in any parametrization. However, whether the fixed point satisfies (\ref{critical_point_moon}) or (\ref{critical_point_star}) might depend on the parametrization.

Linearizing Eq.~(\ref{skalaarvalja_kosmo_vorrand_invariantidena}) around a fixed point $\left( \mathcal{I}_3\left(\Phi_0\right) = 0 , \frac{d}{d\hat{t}}\mathcal{I}_3 = 0 \right)$ gives
\be
\label{linearized_scalar_field_equation_in_terms_of_invariants}
\frac{d}{d\hat{t}} \left( \frac{d}{d\hat{t}}\mathcal{I}_3 \right)  =  - \varepsilon \left.\sqrt{ \frac{3}{\ell^{2}}\mathcal{I}_2}\right|_{\Phi_0}\cdot \frac{d}{d\hat{t}}\mathcal{I}_3 - \left.\frac{1}{2\ell^2}\frac{d^2\mathcal{I}_2}{d\mathcal{I}_3^2} \right|_{\Phi_0} \cdot \mathcal{I}_3 \,,
\ee
or written as a dynamical system 
 \renewcommand{\arraystretch}{2}
\be
\label{Dynamical_system_in_terms_of_invariants}
\begin{pmatrix}
	\frac{d\phantom{t}}{\rule{0ex}{1.5ex}d\hat{t}}\mathcal{I}_3 \\
	\frac{d\phantom{t}}{\rule{0ex}{1.5ex}d\hat{t}}\Pi
\end{pmatrix} =
\left[ 
\begin{matrix}
	0 & 1 \\
	- \frac{1}{\rule{0ex}{1.5ex}2\ell^2}\frac{d^2\mathcal{I}_2}{\rule{0ex}{1.5ex}d\mathcal{I}_3^2} & - \varepsilon \sqrt{ \frac{3}{\rule{0ex}{1.5ex}\ell^{2}}\mathcal{I}_2}
\end{matrix}
\right]_{\Phi_0}
\begin{pmatrix}
		\mathcal{I}_3 \\
		\Pi
\end{pmatrix} \,,
\ee
\renewcommand{\arraystretch}{1}
where $\Pi \equiv \frac{d\phantom{t}}{d\hat{t}}\mathcal{I}_3 $.

\subsection{Solution to the linearized equation}
Solutions of the linearized equation (\ref{linearized_scalar_field_equation_in_terms_of_invariants}) are determined by the eigenvalues of the matrix in Eq.~(\ref{Dynamical_system_in_terms_of_invariants}).
A straightforward calculation shows that the eigenvalues are 
\be
\lambda^{\varepsilon}_\pm = \frac{1}{2\ell}\left[ -\varepsilon \sqrt{3\mathcal{I}_2} \pm \sqrt{3\mathcal{I}_2 - 2 \frac{d^2\mathcal{I}_2}{d\mathcal{I}_3^2}}\right]_{\Phi_0} \,.
\ee
It is clear that these eigenvalues are invariant. As the properties of a fixed point, i.e.~the characteristic features of the solutions near that point, are determined by the real and imaginary parts of the eigenvalues, we can infer that if a fixed point is an attractor in one parametrization, it will be an attractor in any parametrization, etc. The qualitative features of the solutions like convergence and periods of oscillation are independent of the parametrization. 
Writing the eigenvalues in terms of the arbitrary functions $\left\lbrace \mathcal{A}\,, \mathcal{B}\,,\mathcal{V} \right\rbrace$ gives
\beq
\lambda^{\varepsilon}_\pm &=& \frac{1}{\ell\sqrt{\mathcal{A}(\Phi_0)}} \left[ - \varepsilon\sqrt{\frac{3\mathcal{V}}{4\mathcal{A}}} 
\label{eigenvalues_in_terms_of_free_functions}
\pm\sqrt{\frac{3\mathcal{V}}{4\mathcal{A}} - 2\frac{ \left(\mathcal{V}^\prime \mathcal{A} - 2\mathcal{V}\mathcal{A}^\prime \right)^\prime }{2\mathcal{A}\mathcal{B} + 3\left( \mathcal{A}^\prime \right)^2} - \left( \frac{1}{2\mathcal{A}\mathcal{B} + 3\left( \mathcal{A}^\prime \right)^2} \right)^\prime \left(\mathcal{V}^\prime \mathcal{A} - 2\mathcal{V}\mathcal{A}^\prime\right)  } \right]_{\Phi_0} .
\eeq
Here under the second square root we have realized that
if $\frac{\mathcal{I}_2^\prime}{\mathcal{I}_3^\prime}=0$ then also $\frac{\mathcal{I}_2^\prime}{\left(\mathcal{I}_3^\prime\right)^2}=0$
due to the assumption (\ref{nonvanishing_F}).
From the eigenvalues (\ref{eigenvalues_in_terms_of_free_functions}) we see that if $\frac{1}{\mathcal{F}} \equiv \frac{4\mathcal{A}^2}{2\mathcal{A}\mathcal{B} + 3\left( \mathcal{A}^\prime \right)^2} = 0$ and $\left(\frac{1}{\mathcal{F}}\right)^\prime \equiv \left( \frac{4\mathcal{A}^2}{2\mathcal{A}\mathcal{B} + 3\left( \mathcal{A}^\prime \right)^2} \right)^\prime = 0$ at the same value $\Phi_0$, then one of the eigenvalues is zero, hence its real part is also zero and the fixed point is nonhyperbolic. Therefore the assumptions (\ref{nonvanishing_F})-(\ref{non_vanishing_first_derivative_of_1_over_F}) are necessary conditions for studying the properties of the fixed points by using linearization.

If the eigenvalues are different then the general solution for equation (\ref{linearized_scalar_field_equation_in_terms_of_invariants}) reads
\be
\label{solution_for_linearized_equation_in_terms_of_invariants}
\mathcal{I}_3(\hat{t}) = M_1 e^{\lambda^{\varepsilon}_{+}\hat{t}} + M_2 e^{\lambda^{\varepsilon}_{-}\hat{t}} \,,
\ee
where $M_1$ and $M_2$ are constants of integration.
We can make use of the Taylor expansion (\ref{Taylor_series_of_Phi}) to write out the solution for scalar field $\Phi$ from  (\ref{solution_for_linearized_equation_in_terms_of_invariants}).
If the scalar field value at that fixed point is determined by the condition (\ref{critical_point_moon}), then the leading term in the Taylor expansion is linear and gives
\be
\Phi(\hat{t}) - \Phi_{\bullet} \approx 
\pm \left. \frac{1}{\sqrt{\mathcal{F}}} \right|_{\Phi_\bullet} \mathcal{I}_3(\hat{t})\,.
\ee
On the other hand, if the scalar field value is determined by the condition (\ref{critical_point_star}), then the first coefficient of the Taylor expansion (\ref{Taylor_series_of_Phi}) vanishes and the leading term is of the second order (\ref{non_crossing_condition}),
\be
\Phi(\hat{t}) - \Phi_{\star} \approx 
0 + \left.\frac{1}{4}\left(\frac{1}{\mathcal{F}}\right)^\prime\right|_{\Phi_\star} \cdot \mathcal{I}_3^2(\hat{t})\,.
\ee
In the latter case the solution is
\be
\Phi(\hat{t}) - \Phi_{\star} \approx \left.\frac{1}{4}\left(\frac{1}{\mathcal{F}}\right)^\prime\right|_{\Phi_\star} \left(M_1 e^{\lambda^{\varepsilon}_{+}\hat{t}} + M_2 e^{\lambda^{\varepsilon}_{-}\hat{t}}\right)^2 \,.
\ee
Here the underlying perturbed equation for $\Phi$ could not have been  linear one and this is exactly in accord with the approach in Ref.~\cite{JKS2010}. See also the discussion around Eq.~(\ref{non_crossing_condition}).

The redefinition of time $\hat{t} \to t$ 
should rigorously be given as an integral due to Eq.~(\ref{coordinate_transformation_t_to_t_hat}). Since $\mathcal{A}$ is assumed to be always positive, nondiverging and nonvanishing we conclude that we can just substitute $\hat{t}=\sqrt{\mathcal{A}}t$ because this has no effect on the properties of the fixed point.

Analysis of the $\lambda^{\varepsilon}_{+} = \lambda^{\varepsilon}_{-}$ case can be handled in a similar manner.

\subsection{Eigenvalues in different parametrizations}
Writing the eigenvalues in the general parametrization (\ref{eigenvalues_in_terms_of_free_functions}) in terms of the Jordan frame BDBW parametrization gives
\be
\lambda_\pm^{\varepsilon \,(BDBW)}
= \frac{1}{\ell\sqrt{\Psi_0}} \left[ -\varepsilon \sqrt{\frac{3\mathcal{V}}{4\Psi}} \pm\sqrt{\frac{3\mathcal{V}}{4\Psi} - 2\frac{ \left(\mathcal{V}^\prime \Psi - 2\mathcal{V} \right)^\prime }{2\omega + 3} - \left( \frac{1}{2\omega+3} \right)^\prime \left(\mathcal{V}^\prime \Psi - 2\mathcal{V}\right)  } \right]_{\Psi_0} \,.
\ee
For the more usual fixed point at $\Phi_\bullet$ this result coincides with the eigenvalues found in Refs.~\cite{Faraoni_Jensen_Theuerkauf, JKS2008}, while for the nonlinear situation of $\Phi_\star$ this result matches the solutions obtained in Ref.~\cite{JKS2010b}.
The eigenvalues (\ref{eigenvalues_in_terms_of_free_functions}) expressed in the JF BEPS parametrization read
\be
\lambda_\pm^{\varepsilon(BEPS)} = \frac{1}{\ell\sqrt{F(\phi_0)}}\left[ -\varepsilon \sqrt{ \frac{3\mathcal{V}}{4F} } \pm \sqrt{ \frac{3\mathcal{V}}{4F} - 2\frac{ \mathcal{V}^{\prime\prime} F^2 - 2\mathcal{V} \left( \left(F^\prime\right)^2 + F\,F^{\prime\prime} \right) }{F \left(2F + 3\left(F^\prime\right)^2 \right)} }\right]_{\phi_0} \,.
\ee
For instance these can be compared to the present accelerating epoch in the model with specific curvature coupling function but general potential in Ref.~\cite{Hrycyna}. The fixed point stability condition is determined by the real part of the eigenvalues. Note that in the JF BEPS parametrization only the $\Phi_\bullet$ case (\ref{critical_point_moon}) can be realized. The last remark holds true also for the Einstein frame canonical parametrization for which the eigenvalues
\be
\lambda_\pm^{\varepsilon \,(EF\,can.)} =  \frac{1}{2\ell}\left[ -\varepsilon \sqrt{3\mathcal{V}} \pm\sqrt{ 3\mathcal{V} - 2\mathcal{V}^{\prime\prime} } \right]_{\varphi_0} \,
\ee
obtained from (\ref{eigenvalues_in_terms_of_free_functions}) are in accord with the results for the general potential case analyzed in Ref.~\cite{Leon}, as well as the solutions in Ref.~\cite{Damour_Nordtvedt}.


\vspace{-3mm}

\section{Conclusion}

We have considered general scalar-tensor gravity without derivative couplings. Using the transformation properties of four arbitrary functions $\left\lbrace \mathcal{A}(\Phi)\,,\mathcal{B}(\Phi)\,,\mathcal{V}(\Phi)\,,\alpha(\Phi) \right\rbrace$ we have constructed three functions $\mathcal{I}_1$,
$\mathcal{I}_2$, $\mathcal{I}_3$ of the scalar field $\Phi$ that are invariant 
under a local rescaling of the metric tensor and the scalar field reparametrization. These three invariants can be used to define infinitely many analogical invariants via three procedures: i) forming arbitrary functions of these; 
ii) introducing quotient of derivatives $\mathcal{I}_m\equiv\frac{\mathcal{I}_k^\prime}{\mathcal{I}_l^\prime}$; iii) integrating in the sense of indefinite integral $\mathcal{I}_r \equiv \int \mathcal{I}_n\mathcal{I}_p' d\Phi$. Using these invariants we have written down the rules that easily allow to transform invariant quantities from three distinct parametrizations (JF BDBW, JF BEPS and EF canonical) into the general one. 
Useful formulas are gathered into Table \ref{Tabel_1}. 
By introducing an invariant object $\hat{g}_{\mu\nu} \equiv \mathcal{A}g_{\mu\nu}$ we can write the equations of motion and the action in terms of invariants.

We argue that physical observables appear as invariant quantities.
This is illustrated by PPN parameters and the features of cosmological solutions near scalar field fixed points.
We demonstrate that these invariant expressions accommodate the results obtained in earlier literature for distinct conformal frames and reparametrizations of the scalar field.
In a particular case this formalism provides a nice explanation to the correspondence of linear and nonlinear approximate solutions in the Einstein and Jordan frames.

As an outlook it would be interesting to see, whether the invariant variables proposed here would help to clarify the contested issues of 
the frame dependence of cosmological perturbations and quantum corrections in STG. 
As an extension one may consider whether an analogous reasoning can be carried out for more general scalar-tensor theories of gravity with derivative couplings 
and higher order derivatives in action \cite{Horndeski}, where the role of conformal transformation seems to be taken over by disformal transformation
\cite{disformal}.


\medskip
{\bf Acknowledgements}

This work was supported by the Estonian Science Foundation
Grant No. 8837, by the Estonian Ministry for Education and Science
Institutional Research Support Project IUT02-27 and by the European Union through 
the European Regional Development Fund (Centre of Excellence TK114).



\begin{thebibliography}{99}

\bibitem{jordanfierz}
	P.~Jordan,
	Naturwiss. {\bf 26}, 417 (1938);
	{\it Schwerkraft und Weltall, Grundlagen der Theoretische Kosmologie}, 
	(Vieweg und Sohns, Braunschweig, 1952);
	Z. Phys. {\bf 157}, 112 (1959);
	M.~Fierz, Helv.\ Phys.\ Acta {\bf 29}, 128 (1956).

\bibitem{BDBW}
	C.~Brans and R.~H.~Dicke,
	Phys.\ Rev.\  {\bf 124}, 925  (1961);
	%
	P.~G.~Bergmann,
	Int.\ J.\ Theor.\ Phys.\  {\bf 1}, 25 (1968);
	R.~V.~Wagoner,
	Phys.\ Rev.\  D {\bf 1}, 3209 (1970).


\bibitem{stg_books}
	Y. ~Fujii and K. ~Maeda,
	{\it The scalar-tensor theory of gravitation},
	(Cambridge University Press, Cambridge, 2003);
	%
	V. ~Faraoni,
	{\it  Cosmology in scalar-tensor gravity},
	(Kluwer Academic Publishers, Dordrecht, 2004).

\bibitem{Flanagan} 
\'E.~\'E.~Flanagan, 
Class.\ Quant.\ Grav.\  {\bf 21}, 3817 (2004)
\href{http://arxiv.org/abs/gr-qc/0403063}{[gr-qc/0403063]}.

\bibitem{deser}
	S.~Deser,
    Annals Phys.\  {\bf 59}, 248 (1970);
    %
    I.~Bars, P.~Steinhardt and N.~Turok, 
    Phys.\ Rev.\ D {\bf 89}, 043515 (2014) \href{http://arxiv.org/abs/1307.1848}{[arXiv:1307.1848 [hep-th]]};
%
    A.~Padilla, D.~Stefanyszyn and M.~Tsoukalas, 
    Phys.\ Rev.\ D {\bf 89}, 065009 (2014) \href{http://arxiv.org/abs/1312.0975}{[arXiv:1312.0975 [hep-th]]}. 

	\bibitem{Horndeski}
	G.~W.~Horndeski,
	Int.\ J.\ Theor.\ Phys.\  {\bf 10}, 363 (1974);
	%
	C.~Deffayet, S.~Deser and G.~Esposito-Far\`{e}se,
	Phys.\ Rev.\ D {\bf 80}, 064015 (2009)
	\href{http://arxiv.org/abs/0906.1967}{[arXiv:0906.1967 [gr-qc]]};
	%
	C.~Deffayet, X.~Gao, D.~A.~Steer and G.~Zahariade,
	Phys.\ Rev.\ D {\bf 84}, 064039 (2011)
	\href{http://arxiv.org/abs/1103.3260}{[arXiv:1103.3260 [hep-th]]};
	%
	J.~Gleyzes, D.~Langlois, F.~Piazza and F.~Vernizzi,
	\href{http://arxiv.org/abs/1404.6495}{[arXiv:1404.6495 [hep-th]]}.
	%
	X.~Gao,
    Phys.\ Rev.\ D {\bf 90}, 081501 (2014) \href{http://arxiv.org/abs/1406.0822}{[arXiv:1406.0822 [gr-qc]]};
    %
    X.~Gao,
    Phys.\ Rev.\ D {\bf 90}, 104033 (2014) \href{http://arxiv.org/abs/1409.6708}{[arXiv:1409.6708 [gr-qc]]}.

	\bibitem{dark_energy}
	E.~J.~Copeland, M.~Sami and S.~Tsujikawa,
	Int.\ J.\ Mod.\ Phys.\ D {\bf 15}, 1753 (2006)
	\href{http://arxiv.org/abs/hep-th/0603057}{[hep-th/0603057]};
	%
	S.~Tsujikawa,
	Lect.\ Notes Phys.\  {\bf 800}, 99 (2010)
	\href{http://arxiv.org/abs/1101.0191}{[arXiv:1101.0191 [gr-qc]]};
	%
	T.~Clifton, P.~G.~Ferreira, A.~Padilla and C.~Skordis,
	Phys.\ Rept.\  {\bf 513}, 1 (2012)
	\href{http://arxiv.org/abs/1106.2476}{[arXiv:1106.2476 [astro-ph.CO]]}.

	\bibitem{Higgs_inflation}
	F.~L.~Bezrukov and M.~Shaposhnikov,
	Phys.\ Lett.\ B {\bf 659}, 703 (2008)
	\href{http://arxiv.org/abs/0710.3755}{[arXiv:0710.3755 [hep-th]]};
	%
	F.~Bezrukov,
	Class.\ Quant.\ Grav.\  {\bf 30}, 214001 (2013)
	\href{http://arxiv.org/abs/1307.0708}{[arXiv:1307.0708 [hep-ph]]};
	%
	C.~F.~Steinwachs,
	{\it Non-minimal Higgs Inflation and Frame Dependence in Cosmology},
	(Springer, 2014);
	%
	K.~Kamada, T.~Kobayashi, T.~Takahashi, M.~Yamaguchi and J.~Yokoyama,
	Phys.\ Rev.\ D {\bf 86}, 023504 (2012)
	\href{http://arxiv.org/abs/1203.4059}{[arXiv:1203.4059 [hep-ph]]}.

\bibitem{dicke}
	R.~H.~Dicke, Phys.\ Rev.\ {\bf 125}, 2163 (1962).

\bibitem{Faraoni:1998qx}
V.~Faraoni, E.~Gunzig and P.~Nardone,
Fund.\ Cosmic Phys.\  {\bf 20}, 121 (1999)
\href{http://arxiv.org/abs/gr-qc/9811047}{[gr-qc/9811047]}.


\bibitem{physical_equivalence}
K.~Nozari and S.~Davood Sadatian,
Mod.\ Phys.\ Lett.\ A {\bf 24}, 3143 (2009)
\href{http://arxiv.org/abs/0905.0241}{[arXiv:0905.0241 [gr-qc]]};
%
A.~Bhadra, K.~Sarkar, D.~P.~Datta and K.~K.~Nandi,
Mod.\ Phys.\ Lett.\ A {\bf 22}, 367 (2007)
\href{http://arxiv.org/abs/gr-qc/0605109}{[gr-qc/0605109]};
%
S.~Capozziello, P.~Martin-Moruno and C.~Rubano,
Phys.\ Lett.\ B {\bf 689}, 117 (2010)
\href{http://arxiv.org/abs/1003.5394}{[arXiv:1003.5394 [gr-qc]]};
%
C.~Corda,
Astropart.\ Phys.\  {\bf 34}, 412 (2011)
\href{http://arxiv.org/abs/1010.2086}{[arXiv:1010.2086 [gr-qc]]};
%
A.~Stabile, A.~Stabile and S.~Capozziello,
Phys.\ Rev.\ D {\bf 88}, 124011 (2013)
\href{http://arxiv.org/abs/1310.7097}{[arXiv:1310.7097 [gr-qc]]};
%
Y.~N.~Obukhov and D.~Puetzfeld,
\href{http://arxiv.org/abs/1404.6977}{[arXiv:1404.6977 [gr-qc]]}.

\bibitem{faraoni_units}
V.~Faraoni and S.~Nadeau,
Phys.\ Rev.\ D {\bf 75}, 023501 (2007)
\href{http://arxiv.org/abs/gr-qc/0612075}{[gr-qc/0612075]}.

\bibitem{quiros}
I.~Quiros, R.~Garcia-Salcedo, J.~E.~M.~Aguilar and T.~Matos,
Gen.\ Rel.\ Grav.\  {\bf 45}, 489 (2013)
\href{http://arxiv.org/abs/1108.5857}{[arXiv:1108.5857 [gr-qc]]};
    %
    E.~Scholz,
    \href{http://arxiv.org/abs/1111.3220}{[arXiv:1111.3220 [math.HO]]}.

\bibitem{JKS_Einstein_Jordan} 
L.~J\"arv, P.~Kuusk and M.~Saal,
Phys.\ Rev.\ D {\bf 76}, 103506 (2007)
\href{http://arxiv.org/abs/0705.4644}{[arXiv:0705.4644 [gr-qc]]}.

\bibitem{cosmological_perturbations}
J.~Weenink and T.~Prokopec,
Phys.\ Rev.\ D {\bf 82}, 123510 (2010)
\href{http://arxiv.org/abs/1007.2133}{[arXiv:1007.2133 [hep-th]]};
%
J.~O.~Gong, J.~c.~Hwang, W.~I.~Park, M.~Sasaki and Y.~S.~Song,
JCAP {\bf 1109}, 023 (2011)
\href{http://arxiv.org/abs/1107.1840}{[arXiv:1107.1840 [gr-qc]]};
%
T.~Kubota, N.~Misumi, W.~Naylor and N.~Okuda,
JCAP {\bf 1202}, 034 (2012)
\href{http://arxiv.org/abs/1112.5233}{[arXiv:1112.5233 [gr-qc]]};
%
T.~Prokopec and J.~Weenink,
JCAP {\bf 1212}, 031 (2012)
\href{http://arxiv.org/abs/1209.1701}{[arXiv:1209.1701 [gr-qc]]};
%
T.~Prokopec and J.~Weenink,
JCAP {\bf 1309}, 027 (2013)
\href{http://arxiv.org/abs/1304.6737}{[arXiv:1304.6737 [gr-qc]]};
%
T.~Chiba and M.~Yamaguchi,
JCAP {\bf 1310}, 040 (2013)
\href{http://arxiv.org/abs/1308.1142}{[arXiv:1308.1142 [gr-qc]]};
	%
    J.~White, M.~Minamitsuji and M.~Sasaki,
    JCAP {\bf 1207}, 039 (2012) \href{http://arxiv.org/abs/1205.0656}{[arXiv:1205.0656 [astro-ph.CO]]};
	%
    J.~White, M.~Minamitsuji and M.~Sasaki,
    JCAP {\bf 1309}, 015 (2013) \href{http://arxiv.org/abs/1306.6186}{[arXiv:1306.6186 [astro-ph.CO]]}. 


\bibitem{quantum_corrections}
M.~Artymowski, Y.~Ma and X.~Zhang,
Phys.\ Rev.\ D {\bf 88}, 104010 (2013)
\href{http://arxiv.org/abs/1309.3045}{[arXiv:1309.3045 [gr-qc]]};
%
D.~P.~George, S.~Mooij and M.~Postma,
JCAP {\bf 1402}, 024 (2014)
\href{http://arxiv.org/abs/1310.2157}{[arXiv:1310.2157 [hep-th]]};
%
J.~Ren, Z.~Z.~Xianyu and H.~J.~He,
JCAP {\bf 1406}, 032 (2014)
\href{http://arxiv.org/abs/1404.4627}{[arXiv:1404.4627 [gr-qc]]}.



	\bibitem{Kamenshchik:2014waa} 
	A.~Y.~Kamenshchik and C.~F.~Steinwachs,
	\href{http://arxiv.org/abs/1408.5769}{[arXiv:1408.5769 [gr-qc]]}.



\bibitem{catena}
R.~Catena, M.~Pietroni and L.~Scarabello,
Phys.\ Rev.\ D {\bf 76}, 084039 (2007)
\href{http://arxiv.org/abs/astro-ph/0604492}{[astro-ph/0604492]};
%
R.~Catena, M.~Pietroni and L.~Scarabello,
J.\ Phys.\ A {\bf 40}, 6883 (2007)
\href{http://arxiv.org/abs/hep-th/0610292}{[hep-th/0610292]}.


	\bibitem{postma}
	M.~Postma and M.~Volponi,
    Phys.\ Rev.\ D {\bf 90}, no. 10, 103516 (2014) \href{http://arxiv.org/abs/1407.6874}{[arXiv:1407.6874 [astro-ph.CO]]}.


\bibitem{JKSV_1}
  L.~J\"arv, P.~Kuusk, M.~Saal and O.~Vilson,
  J.\ Phys.\ Conf.\ Ser.\  {\bf 532}, 012011 (2014).

\bibitem{Bronnikov:2002kf} 
  K.~A.~Bronnikov,
  J.\ Math.\ Phys.\  {\bf 43}, 6096 (2002) \href{http://arxiv.org/abs/gr-qc/0204001}{[arXiv:gr-qc/0204001]};

  K.~A.~Bronnikov and A.~A.~Starobinsky,
  JETP Lett.\  {\bf 85}, 1 (2007)
  [Pisma Zh.\ Eksp.\ Teor.\ Fiz.\  {\bf 85}, 3 (2007)] \href{http://arxiv.org/abs/gr-qc/0612032}{[arXiv:gr-qc/0612032]}.

\bibitem{BEPS}
B.~Boisseau, G.~Esposito-Far\`{e}se, D.~Polarski and A.~A.~Starobinsky,
Phys.\ Rev.\ Lett.\  {\bf 85}, 2236 (2000)
\href{http://arxiv.org/abs/gr-qc/0001066}{[gr-qc/0001066]}.
%
G.~Esposito-Far\`{e}se and D.~Polarski,
Phys.\ Rev.\ D {\bf 63}, 063504 (2001)
\href{http://arxiv.org/abs/gr-qc/0009034}{[gr-qc/0009034]}.


\bibitem{Wald}
 R.~M.~Wald,
 {\it General Relativity},
 (The University of Chicago Press, Chicago, 1984).

\bibitem{deruelle_sasaki}
N.~Deruelle and M.~Sasaki, 
  Springer Proc.\ Phys.\ {\bf 137}, 247 (2011) \href{http://arxiv.org/abs/1007.3563}{[arXiv:1007.3563 [gr-qc]]}. 

\bibitem{romero}
  C.~Romero, J.~B.~Fonseca-Neto and M.~L.~Pucheu,
  Class.\ Quant.\ Grav.\ {\bf 29}, 155015 (2012) \href{http://arxiv.org/abs/1201.1469}{[arXiv:1201.1469 [gr-qc]]};
%
  R.~Aguila, J.~E.~Madriz Aguilar, C.~Moreno and M.~Bellini,
  Eur.\ Phys.\ J.\ C {\bf 74}, no. 11, 3158 (2014) \href{http://arxiv.org/abs/1408.4839}{[arXiv:1408.4839 [gr-qc]]}.

\bibitem{ohanlon} 
  J.~O'Hanlon,
  Phys.\ Rev.\ Lett.\  {\bf 29}, 137 (1972).
  
\bibitem{JKS2008}
	L.~J\"arv, P.~Kuusk and M.~Saal,
	Phys.\ Rev.\ D {\bf 78}, 083530 (2008) \href{http://arxiv.org/abs/0810.5038}{[arXiv:0810.5038 [gr-qc]]}.  
  
  
\bibitem{Hohmann:2013rba} 
M.~Hohmann, L.~J\"arv, P.~Kuusk and E.~Randla,
Phys.\ Rev.\ D {\bf 88}, 084054 (2013)
[Erratum-ibid.\ D {\bf 89}, 069901 (2014)]
\href{http://arxiv.org/abs/1309.0031}{[arXiv:1309.0031 [gr-qc]]}.

\bibitem{ppn_bdbw}
K.~Nordtvedt, Jr.,
Astrophys.\ J.\  {\bf 161}, 1059 (1970);
%
G.~J.~Olmo,
Phys.\ Rev.\ D {\bf 72}, 083505 (2005)
\href{http://arxiv.org/abs/gr-qc/0505135}{[gr-qc/0505135]};
%
L.~Perivolaropoulos,
Phys.\ Rev.\ D {\bf 81}, 047501 (2010)
\href{http://arxiv.org/abs/0911.3401}{[arXiv:0911.3401 [gr-qc]]}.

\bibitem{ppn_ef}
  T.~Damour and G.~Esposito-Far\`{e}se,
  Class.\ Quant.\ Grav.\  {\bf 9}, 2093 (1992).

\bibitem{scharer}
  A.~Sch\"arer, R.~Ang\'elil, R.~Bondarescu, P.~Jetzer and A.~Lundgren,
  Phys.\ Rev.\ D. {\bf 90}, 123005 (2014)
  \href{http://arxiv.org/abs/1410.7914}{[arXiv:1410.7914 [gr-qc]]}.



\bibitem{JKS2010}
L.~J\"arv, P.~Kuusk and M.~Saal,
Phys.\ Rev.\ D {\bf 81}, 104007 (2010)
\href{http://arxiv.org/abs/1003.1686}{[arXiv:1003.1686 [gr-qc]]}.


\bibitem{Faraoni_Jensen_Theuerkauf}
V.~Faraoni, M.~N.~Jensen and S.~A.~Theuerkauf,
Class.\ Quantum\ Grav. {\bf 23}, 4215 (2006)
\href{http://arxiv.org/abs/gr-qc/0605050v1}{[arXiv:gr-qc/0605050]}.

\bibitem{JKS2010b} 
L.~J\"arv, P.~Kuusk and M.~Saal,
Phys.\ Lett.\ B {\bf 694}, 1 (2010)
\href{http://arxiv.org/abs/1006.1246}{[arXiv:1006.1246 [gr-qc]]}.

\bibitem{Hrycyna}
O.~Hrycyna and M.~Szydlowski,
JCAP {\bf 1012}, 016 (2010)
\href{http://arxiv.org/abs/1008.1432}{[arXiv:1008.1432 [astro-ph.CO]]}; 
O.~Hrycyna and M.~Szydlowski,
JCAP {\bf 0904}, 026 (2009)
\href{http://arxiv.org/abs/0812.5096}{[arXiv:0812.5096 [hep-th]]}.

\bibitem{Leon}
G.~Leon,
Class.\ Quant.\ Grav.\  {\bf 26}, 035008 (2009)
\href{http://arxiv.org/abs/0812.1013}{[arXiv:0812.1013 [gr-qc]]}.

\bibitem{Damour_Nordtvedt}
T.~Damour, K.~Nordtvedt,
Phys.\ Rev.\ D {\bf 48}, 3436 (1993). 




\bibitem{disformal}
D.~Bettoni and S.~Liberati,
Phys.\ Rev.\ D {\bf 88}, 084020 (2013)
\href{http://arxiv.org/abs/1306.6724}{[arXiv:1306.6724 [gr-qc]]};
%
M.~Zumalac\'arregui and J.~Garc\'ia-Bellido,
Phys.\ Rev.\ D {\bf 89}, 064046 (2014)
\href{http://arxiv.org/abs/1308.4685}{[arXiv:1308.4685 [gr-qc]]};
%
M.~Minamitsuji,
Phys.\ Lett.\ B {\bf 737}, 139 (2014)
\href{http://arxiv.org/abs/1409.1566}{[arXiv:1409.1566 [astro-ph.CO]]}. 





\end{thebibliography}
\end{document}